\documentclass[3p,times,a4paper,british]{elsarticle}

\usepackage{amssymb,amsmath,amsthm,amsfonts}
\usepackage[T1]{fontenc}
\usepackage[latin1]{inputenc}
\usepackage{textcomp}
\usepackage{url}
\usepackage{doi}
\usepackage{natbib}
\biboptions{sort&compress}
\usepackage{babel}
\usepackage{graphicx}
\usepackage{adjustbox}
\usepackage{caption}
\usepackage[usenames,dvipsnames,svgnames,table]{xcolor}
\usepackage[font=footnotesize]{subcaption}
\usepackage{microtype}
\usepackage{booktabs}
\usepackage{multirow}
\usepackage{xspace}
\usepackage{todonotes}
\usepackage[version=3]{mhchem}
\usepackage[per-mode=symbol]{siunitx}
\usepackage{tablefootnote}
\usepackage{hyperref}
\usepackage{array}
\usepackage{lineno}

\renewcommand*{\vec}[1]{\boldsymbol{#1}}

\newcommand*{\od}[3][]{\frac{\dif^{#1}#2}{\dif{#3}^{#1}}}
\newcommand*{\pdc}[3]{\left(\frac{\partial #1}{\partial #2}\right)_{#3}}
\newcommand*{\ped}[1]{\ensuremath{_{\text{#1}}}}

\newcommand*{\coto}{\ensuremath{\text{CO}_{\text{2}}}\xspace}
\newcommand*{\nto}{\ensuremath{\text{N}_{\text{2}}}\xspace}
\newcommand*{\hto}{\ensuremath{\text{H}_{\text{2}}}\xspace}
\newcommand*{\htoo}{\ensuremath{\text{H}_{\text{2}}\text{O}}\xspace}
\newcommand*{\oto}{\ensuremath{\text{O}_{\text{2}}}\xspace}

\newcommand*{\chf}{\ensuremath{\text{CH}_{\text{4}}}\xspace}
\newcommand*{\soto}{\ensuremath{\text{SO}_{\text{2}}}\xspace}

\newcommand*{\liq}{\ell}%
\newcommand*{\solid}{\text{s}}%
\newcommand*{\gas}{\text{g}}
\newcommand*{\spec}{\text{spec}}
\newcommand*{\sat}{\text{sat}}

\newcommand*{\tr}{\text{tr}}

\newcommand*{\cfl}{\ensuremath{\text{CFL}}}

\newcommand*{\nondim}[1]{\ensuremath{\mathit{#1}}}
\newcommand*{\Rey}{\nondim{Re}}

\newcommand*{\Mach}{\nondim{Ma}}

\newcommand*{\transp}{\text{T}}
\newcommand*{\rightedge}{\text{R}}
\newcommand*{\leftedge}{\text{L}}

\DeclareSIUnit\atm{atm}

\makeatletter
\newcommand*{\dif}{\@ifnextchar^{\DIfF}{\DIfF^{}}}
\def\DIfF^#1{\mathop{\mathrm{\mathstrut d}}\nolimits^{#1}\gobblesp@ce}
\def\gobblesp@ce{\futurelet\diffarg\opsp@ce}
\def\opsp@ce{%
  \let\DiffSpace\!%
  \ifx\diffarg(%
    \let\DiffSpace\relax
  \else
    \ifx\diffarg[%
      \let\DiffSpace\relax
    \else
      \ifx\diffarg\{%
        \let\DiffSpace\relax
      \fi\fi\fi\DiffSpace}
\makeatother

\graphicspath{{./gfx/}}

\begin{document}

\begin{frontmatter}
\title{Computation of three-dimensional three-phase flow of carbon
  dioxide using a high-order WENO scheme}
\author{Magnus Aa. Gjennestad}
\author{Andrea Gruber}
\author{Karl Yngve Lerv{\aa}g}
\author{{\O}yvind Johansen}
\author{{\AA}smund Ervik}
\author{Morten Hammer}
\author{Svend~Tollak~Munkejord\corref{cor}}
\ead{svend.t.munkejord [a] sintef.no}

\address{SINTEF Energy Research, P.O. Box 4761 Sluppen, NO-7465
  Trondheim, Norway}
\cortext[cor]{Corresponding author.}

\begin{abstract}
  We have developed a high-order numerical method for the 3D simulation of
  viscous and inviscid multiphase flow described by a homogeneous
  equilibrium model and a general equation of state. Here we focus on
  single-phase, two-phase (gas-liquid or gas-solid) and three-phase
  (gas-liquid-solid) flow of \coto whose thermodynamic properties are
  calculated using the Span--Wagner reference equation of state. The
  governing equations are spatially discretized on a uniform Cartesian grid
  using the finite-volume method with a fifth-order weighted essentially
  non-oscillatory (WENO) scheme and the robust first-order centred (FORCE)
  flux. The solution is integrated in time using a third-order
  strong-stability-preserving Runge--Kutta method. We demonstrate close to
  fifth-order convergence for advection-diffusion and for smooth single-
  and two-phase flows. Quantitative agreement with experimental data is
  obtained for a direct numerical simulation of an air jet flowing from a
  rectangular nozzle. Quantitative agreement is also obtained for the shape
  and dimensions of the barrel shock in two highly underexpanded \coto
  jets.
\end{abstract}

\end{frontmatter}

\enlargethispage{-2\baselineskip}

\section{Introduction}
\label{sec:intro}

The deployment of \coto capture and storage (CCS) is regarded as a key
strategy to mitigate global warming \citep{IPCC-syn2014}. To design and
operate CCS systems in a safe and cost-effective way, accurate data and
models are needed \citep{Munkejord2016}. This includes models and methods
to simulate the near field of a \coto jet resulting from the decompression
of equipment containing high-pressure \coto. The data from these near-field
simulations are e.g.\ used as input for less resolved simulations of the
dispersion of \coto in the terrain
\citep{Witlox2014,Woolley2014,Jamois2015}.

This type of scenario puts some requirements on the models and numerical
methods to be used. Depressurization of \coto from supercritical pressures
typically involves complex three-phase (gas-liquid-solid) flow. Describing
this kind of flow necessitates a multiphase flow model and an equation of
state (EOS) that is accurate and capable of capturing the three-phase
behaviour \citep{Hammer2013,Wareing2013}. For high vessel pressures, the
\coto jet resulting from a leak will form a shock, which the numerical
method must be able to capture. In addition, we
would like the numerical method to maintain discrete conservation of mass,
momentum and energy and, due to computational efficiency, to be of
high order in smooth regions of the computational domain, without producing
spurious oscillations in the solution near discontinuities.

\citet{Wareing2013,Wareing2014} and \citet{Woolley2013} studied \coto jets
using a Reynolds-averaged Navier--Stokes model. The flow model
was combined with a composite EOS \citep{Wareing2013} to
describe three-phase \coto flow. The flow model was solved using
a conservative, shock-capturing second-order scheme, as described by
\citet{Falle1991}.

However, advances have been made in constructing and implementing
finite-volume, shock capturing and conservative numerical methods of higher
order. \citet{Titarev2004} presented a procedure relying on weighted
essentially non-oscillatory (WENO) interpolation \citep{Jiang1996} and
achieved fifth-order convergence for their smooth and inviscid
two-dimensional isentropic vortex problem. Their scheme was extended to
include interpolation of velocity derivatives and computation of viscous
transport of momentum and dissipation of kinetic energy by
\citet{Coralic2014}. Such a numerical scheme is suitable for execution on
parallel computers by domain decomposition \citep{Petsc-web-page}.

In research on numerical methods for compressible multiphase flow,
the ideal-gas and stiffened-gas EOS
\citep{Menikoff1989,Menikoff2007} are commonly employed, due to their
simplicity and relatively large number of applications. This is true
both for 1D \citep{Saurel2008,Zein2010,Rodio2015} and 3D models
\citep[e.g.][]{Coralic2014}. The stiffened-gas EOS can be regarded as a
linearization about a reference state. In many cases, however, it is
necessary to consider more adapted EOSs in order to
achieve the necessary accuracy. This often entails a significantly higher
computational complexity. As an example, \citet{Dumbser2013}
presented an unstructured WENO scheme employing a real EOS for water.

For CCS applications, it is often necessary to describe a large
thermodynamic property space, involving multiple phases, for instance for
the depressurization from a transport pipeline operated at a supercritical
pressure around \SI{100}{\bar} down to atmospheric conditions. In these
cases, an accurate EOS is required \citep{Hammer2013}, such as the one by
\citet{Span1996} (SW). Therefore, in order to perform high-fidelity
near-field studies of \coto jets, we need to combine a high-order numerical
scheme with a general EOS.

This combination would also benefit the development of predictive
fluid-structure models aiding in the design of \coto-transport pipelines
against running fractures \citep{Aursand2016a,Talemi2016}. For practical
and computational reasons, the \coto flow is commonly described using a 1D
model, which implies a simplified description of the pressure forces on the
opening pipe flanks \citep{Aursand2016a}. A full 3D description of the flow
might provide more accurate predictions.

In the present work we want to study complex \coto flows which may be
single phase, two-phase (gas-liquid or gas-solid) or three-phase
(gas-liquid-solid). In doing so, we extend the high-order scheme of
\citet{Titarev2004} and \citet{Coralic2014}, applying it to the homogeneous
equilibrium multiphase flow model and a formulation allowing the use of a
general EOS.  Since the applications we are interested in typically involve
sharp temperature gradients, we include heat conduction in our model and in
the numerical treatment of diffusive fluxes, as well as a
temperature-dependent viscosity.

We validate the implementation of the model and numerical
methods through several test cases, including a turbulent air jet from a
rectangular nozzle. We also demonstrate that the numerical methods exhibit
high-order convergence when dealing with diffusive fluxes and two-phase
flows. Finally, we perform detailed simulations of \coto jets,
employing the SW EOS.

The rest of this paper is organized as follows. Section~\ref{sec:model}
reviews the governing equations, while the treatment of inflow and open
boundary conditions is briefly described in
Section~\ref{sec:bc}. Section~\ref{sec:nummeth} deals with the numerical
methods. Section~\ref{sec:valid} demonstrates the accuracy and robustness
of the scheme, including the direct numerical simulation of an air jet,
while Section~\ref{sec:co2jet} discusses the simulation of a \coto
jet. Section~\ref{sec:concl} concludes the study.

\section{Models}
\label{sec:model}

\subsection{Fluid dynamics}
We consider a three-dimensional flow of a fluid that may consist of
multiple phases. The different phases are assumed to be in local
equilibrium and to move with the same velocity. The flow may then be
described by a homogenous equilibrium model (HEM), which can be
formulated as a system of balance equations,
\begin{equation}
  \label{eq:hem}
  \partial_t \vec{Q} + \partial_x \vec{F}
  + \partial_y \vec{G}
  + \partial_z \vec{H} = \vec{S}(\vec{Q}).
\end{equation}
Here $\vec{F}$, $\vec{G}$ and $\vec{H}$ are the fluxes in the $x$-,
$y$- and $z$-direction, respectively, and $\vec{S}(\vec{Q})$ is the vector
of source terms. The vector $\vec{Q}$ contains the
state variables,
\begin{equation}
  \vec{Q} = \left[ \rho, \rho u_x, \rho u_y, \rho u_z, E \right]^\transp,
\end{equation}
where $\rho$ is the fluid density, $u_x$, $u_y$ and $u_z$ are the flow
velocities and $E$ is the total energy density. Thus the system
\eqref{eq:hem} describes conservation of mass and balance of momentum and
energy of the fluid. The total energy is
\begin{equation}
  \label{eq:total_energy}
  E = \rho e + \frac{1}{2} \rho \left( u_x^2 + u_y^2 + u_z^2 \right),
\end{equation}
where $e$ is the specific internal energy of the fluid. The total
energy is thus the sum of internal and kinetic energy.

The fluxes are
\begin{equation}
  \label{eq:fluxes}
  \vec{F} =
  \begin{bmatrix}
    \rho u_x \\
    \rho u_x^2 + p - \sigma_{xx} \\
    \rho u_y u_x - \sigma_{xy} \\
    \rho u_z u_x - \sigma_{xz} \\
    (E + p) u_x - u_i\sigma_{xi} - \kappa \partial_x T
  \end{bmatrix},
\end{equation}
\begin{equation}
  \vec{G} =
  \begin{bmatrix}
    \rho u_y \\
    \rho u_x u_y - \sigma_{yx} \\
    \rho u_y^2 + p - \sigma_{yy} \\
    \rho u_z u_y - \sigma_{yz} \\
    (E + p) u_y - u_i\sigma_{yi} - \kappa \partial_y T
  \end{bmatrix},
\end{equation}
and
\begin{equation}
  \vec{H} =
  \begin{bmatrix}
    \rho u_z \\
    \rho u_x u_z - \sigma_{zx} \\
    \rho u_y u_z - \sigma_{zy} \\
    \rho u_z^2 + p - \sigma_{zz} \\
    (E + p) u_z - u_i\sigma_{zi} - \kappa \partial_z T
  \end{bmatrix}.
\end{equation}
Herein, $p$ is the fluid pressure, $T$ is the temperature, $\kappa$ is the
thermal conductivity and $\sigma_{ij}$ is the viscous stress tensor. With
these fluxes, and no source terms, \eqref{eq:hem} corresponds to the Euler
equations with added diffusive fluxes for viscous transport of momentum and
conductive transport of heat.

We assume zero bulk viscosity, in which case the viscous
stress tensor is given by the velocity derivatives and the dynamic
viscosity $\eta$ as \cite{Landau1987}
\begin{equation}
  \sigma_{ij} = \eta \left( \partial_i u_j + \partial_j u_i
  - \frac{2}{3} \delta_{ij} \partial_k u_k \right).
\end{equation}

\subsection{Thermophysical properties}
\label{subsec:thermphys}
In order to close the system \eqref{eq:hem}, we must employ a thermodynamic
equation of state (EOS). We assume local thermodynamic phase equilibrium and
consider only pure components. In this paper, we make use of the ideal gas
EOS (IG), the Peng--Robinson \cite{Peng1976} EOS (PR) and the
multi-parameter Span--Wagner \cite{Span1996} reference EOS (SW) for
\coto. Both PR and SW describe gas-liquid systems. By coupling SW to an
additional model for the solid \coto phase, it can be extended to systems
including a solid phase and be used to describe solid formation, as
described in \citet{Hammer2013}.

The thermal conductivity $\kappa$ is assumed constant throughout this
work. The dynamic viscosity $\eta$, however, has a strong
temperature dependence and cannot always be assumed
constant. Therefore, we use the TRAPP extended corresponding state
model due to \citet{Ely1981} for the dynamic viscosity in cases with
large temperature variations.

\section{Boundary conditions}
\label{sec:bc}

\subsection{Nozzle inflow}
\label{sec:bernoulli}
For the \coto jet to be studied, we model inflow through a nozzle
located at the domain boundary.  Boundary conditions in the nozzle
region are set by the isentropic steady-state Bernoulli
equations,
\begin{linenomath}
\begin{align}
  \dif h + u \dif u &= 0, \\
  \dif s &= 0,
\end{align}
\end{linenomath}
for the specific enthalpy $h$, specific entropy $s$ and
velocity $u$ of the fluid. Integrating these to the boundary from some
known rest state behind the nozzle, specified e.g.\ by $T_\infty$ and
$ p_\infty$, we get
\begin{linenomath}
\begin{align}
  s_\infty &= s_{\text{b}}, \\
  h_\infty &= h(s_{\text{b}}, p_{\text{b}}) + \frac{1}{2} u_{\text{b}}^2,
\end{align}
\end{linenomath}
where the boundary values have a subscript $\text{b}$. By setting
$u_{\text{b}}$ equal to the speed of sound and solving the integrated
Bernoulli equations for $p_{\text{b}}$, we obtain the choke pressure
at the nozzle. This procedure thus gives the pressure, entropy and
flow velocity at the boundary and the boundary condition is completely
specified.

\subsection{Non-reflecting boundaries}
\label{sec:openbc}
Many practical flows of interest are located in physical domains that are
unbounded in one or more spatial directions and require the specification
of an artificial boundary in order to make the computational domain
finite. The artificial boundary represents a connection between the
computational domain and the surrounding far field. Care must be taken in
the definition of this \textit{open boundary}. Under-specification or
over-specification of physical boundary conditions would lead to an
ill-posed problem and are a classical cause of numerical instability. In
fluid flows, information about the flow conditions is transmitted across
the open boundaries by physical waves. These open boundaries should allow
waves (especially pressure waves or acoustic waves) to travel freely in and
out of the computational domain. However, the knowledge about the exterior
can often be unsure or absent and additional modelling or qualified guesses
about these flow conditions may be necessary. In particular, the amplitudes of the
outgoing waves may be used as a starting point for the modelling of the
incoming ones. This approach, named Navier--Stokes Characteristic Boundary
Conditions (NSCBC), is utilized in the present work to specify the open
boundaries of the computational domain, as described in the landmark paper
by \citet{Poinsot92} and later refined by \citet{Sutherland03} for the general
context of single-phase, multi-component and reactive flows.

\section{Numerical methods}
\label{sec:nummeth}

The fluid-dynamical model is integrated in time using the
finite-volume method on a uniform Cartesian grid. This method transforms the
coupled system of PDEs \eqref{eq:hem} into a system of coupled ODEs
that can be integrated in time with an appropriate Runge--Kutta
method.

\subsection{Spatial discretization}
\label{sec:spatdisc}
The semi-discrete form of the PDE system \eqref{eq:hem} is obtained by
integrating it over the volume of a cell $i,j,k$ and applying the
divergence theorem,
\begin{linenomath}
\begin{align}
  \od{}{t} \vec{Q}_{i,j,k} = \
  &\frac{1}{\Delta x} \left( \vec{F}_{i-1/2,j,k} - \vec{F}_{i+1/2,j,k} \right)
  + \frac{1}{\Delta y} \left( \vec{G}_{i,j-1/2,k} - \vec{G}_{i,j+1/2,k} \right)
  + \frac{1}{\Delta z} \left( \vec{H}_{i,j,k-1/2} - \vec{H}_{i,j,k+1/2} \right).
\end{align}
\end{linenomath}
Herein, we have defined the volume-averaged state variables for the cell
$i,j,k$,
\begin{linenomath}
\begin{align}
  \label{eq:Q_vol}
  \vec{Q}_{i,j,k} &\equiv \frac{1}{\Delta x \Delta y \Delta z}
  \int_{x_{i-1/2}}^{x_{i+1/2}} \int_{y_{j-1/2}}^{y_{j+1/2}}
  \int_{z_{k-1/2}}^{z_{k+1/2}}
  \vec{Q}\left(x,y,z,t\right) \ \dif x \dif y \dif z,
\end{align}
\end{linenomath}
and the area-averaged fluxes over the cell edges,
\begin{linenomath}
\begin{align}
  \label{eq:F_surf}
  \vec{F}_{i-1/2,j,k} &\equiv \frac{1}{\Delta y \Delta z}
  \int_{y_{j-1/2}}^{y_{j+1/2}} \int_{z_{k-1/2}}^{z_{k+1/2}}
  \vec{F} \left(x_{i-1/2},y,z,t\right) \ \dif y \dif z, \\
  \label{eq:G_surf}
  \vec{G}_{i,j-1/2,k} &\equiv \frac{1}{\Delta x \Delta z}
  \int_{x_{i-1/2}}^{x_{i+1/2}} \int_{z_{k-1/2}}^{z_{k+1/2}}
  \vec{G} \left(x,y_{j-1/2},z,t\right) \ \dif x \dif z, \\
  \label{eq:H_surf}
  \vec{H}_{i,j,k-1/2} &\equiv \frac{1}{\Delta x \Delta y}
  \int_{x_{i-1/2}}^{x_{i+1/2}} \int_{y_{j-1/2}}^{y_{j+1/2}}
  \vec{H} \left(x,y,z_{k-1/2},t\right) \ \dif x \dif y.
\end{align}
\end{linenomath}

Approximating the flux integrals \eqref{eq:F_surf}-\eqref{eq:H_surf} using
one quadrature point per cell edge, one may derive numerical schemes that
are at most second-order. If one instead evaluates the flux integrals using
multiple quadrature points on each cell edge, numerical methods of
higher order can be constructed. The evaluation of the flux integrals is
then done by first computing the numerical flux at each quadrature point,
and then taking some linear combination of the computed fluxes. This
procedure requires reconstruction of the fluid state and derivatives of
velocity and temperature to both sides of the cell edges at each quadrature
point. It also requires high-order numerical volume integration when
calculating the volume-averaged primitive variables $\vec{V}_{i,j,k}$ from
the state variables $\vec{Q}$, as noted by
\citet{Coralic2014}. That is, the integral
\begin{linenomath}
\begin{align}
  \label{eq:V_vol}
  \vec{V}_{i,j,k} &\equiv \frac{1}{\Delta x \Delta y \Delta z}
  \int_{x_{i-1/2}}^{x_{i+1/2}} \int_{y_{j-1/2}}^{y_{j+1/2}}
  \int_{z_{k-1/2}}^{z_{k+1/2}}
  \vec{V}\left(\vec{Q}\left(x,y,z,t\right)\right) \ \dif x \dif y \dif z,
\end{align}
\end{linenomath}
must be approximated numerically using multiple quadrature points per cell
volume and thus the state variables must be reconstructed to these
quadrature points.

\citet{Titarev2004} presented a procedure as outlined above using WENO
interpolation for reconstruction of the fluid states. This was
extended to include reconstruction of derivatives and computation of
diffusive fluxes by \citet{Coralic2014}. We will rely on their methods
in this work and the reader is referred to their works for a more
thorough exposition. We shall here employ a fifth-order WENO scheme
for reconstruction of fluid states and use fourth-order Gaussian
quadrature rules, two quadrature points for each cell edge integral
and four quadrature points for each cell volume integral, in all 2D
simulations. In 3D, we shall use four quadrature points
for each cell edge integral and eight quadratures point
for each cell volume integral. As basic
advective numerical flux, we employ the robust first-order centred
(FORCE) scheme \citep{Toro2000}. Regarding the calculation of the WENO
weights, we employ the relations presented in \citep{Titarev2004}.

\subsubsection{Reconstructed variables}
In order for the fluid states at the quadrature points to be
consistent with the EOS, one must choose a set of five
variables\footnote{Five variables for 3D simulations, four for 2D and
  three for 1D.}, interpolate these to the quadrature points and then
use the interpolated values, the EOS and the thermophysical property
models to compute the remaining variables needed to compute the
fluxes. As noted by e.g. \citet{Coralic2014}, the fluid state may be
reconstructed using many different sets of variables, i.e. the choice
of reconstruction variables is not unique. To avoid spurious
oscillations, however, it will often be necessary to reconstruct in
another set of variables than the state variables $\vec{Q}$
\citep[Sec.~14.4.3]{Toro1999}

\citet{Hammer2013} performed reconstruction in flow velocity, density
and internal energy when performing 1D simulations with second-order
MUSCL reconstruction and the Span--Wagner reference EOS for \coto. For
simulations with high-order WENO reconstruction and the ideal gas or
stiffened-gas EOS, \citet{Titarev2004} and \citet{Coralic2014}
performed reconstruction in the local characteristic
variables. \citet{Coralic2014} obtained the local characteristic
variables by multiplying the vector of primitive variables with a locally
frozen transformation matrix. We will follow their procedure when
using the ideal-gas EOS. However, for more advanced EOSs,
we use a more general procedure which is described in \ref{sec:chars}.

\subsection{Temporal integration}
\label{sec:RK}
For time integration, we use the three-step third-order
strong-stability-preserving Runge--Kutta (RK) method \cite[see
e.g.][]{Ketcheson2005}. Our time steps are limited by a
Courant--Friedrichs--Lewy (CFL) criterion for all cases. This is done in a
similar way as in \cite{Titarev2004}. For cases with viscosity and thermal
conductivity, one must in addition consider the time step restriction
imposed by the diffusive fluxes \cite{Coralic2014}. Given a set of fluid
parameters, the latter restriction will be limiting for the time step
length if fine enough grids are used. In practice, however, we found that
the CFL criterion was sufficient to ensure stability for the grids and
fluids considered in this study.

\subsection{Phase equilibrium}
\label{sec:flashes}
When the balance equations \eqref{eq:hem} are advanced in time, the
mass of each component, and the momentum and total energy of the
mixture are updated in every control volume. This allows the
determination of the specific volume ${v}$ and internal energy
${e}$. For given ${e}$ and ${v}$, the equilibrium phase distribution
and the intensive variables temperature $T$ and pressure $p$ must be
determined. This calculation is called a \textit{flash}, or more
specifically an \texorpdfstring{${e}{v}$}{ev}-flash. Mathematically,
the \texorpdfstring{${e}{v}$}{ev}-flash represents a global
maximization of entropy in the temperature-pressure-phase-fraction
space, subject to constraints on mass and internal energy. A
challenging part of this calculation is to determine which phases are
present. Under the assumption of full equilibrium (mechanical, thermal
and chemical), the phases present must have the same pressure,
temperature and chemical potential. For our numerical methods, we
guess which phases are present and then solve to meet the
constrains. During this iterative procedure, all phases present are in
full equilibrium. When the constraints are satisfied for the trial set
of phases, it must be determined if the solution is a local or a
global solution by introducing or removing phases. The phase
distribution maximizing the entropy is the
\texorpdfstring{${e}{v}$}{ev}-flash solution.

At different steps in the model integration, it becomes necessary to
solve different flash problems. These are still global optimization
problems, but they have constraints other than mass and internal energy.
Depending on what information is available at a given step, we solve
one of the following problems to obtain the equilibrium state.
\begin{itemize}
\item Equilibrium calculation with specified internal energy
  ${e}_\spec$ and specific volume ${v}_\spec$
  (${e}{v}$-flash)
\item Equilibrium calculation with specified entropy
  ${s}_\spec$ and specific volume ${v}_\spec$
  (${s}{v}$-flash)
\item Equilibrium calculation with specified entropy
  ${s}_\spec$ and pressure $p_\spec$ ($p{s}$-flash)
\end{itemize}

When performing reconstruction in internal energy, density and velocity, we
have a known internal energy $e_\spec$ and specific volume ${v}_\spec$. To
determine the temperature, pressure and phase distribution, an
\texorpdfstring{${e}{v}$}{ev}-flash must be solved.

To set boundary-condition states from the nozzle inflow model (see
Section~\ref{sec:bernoulli}), we must calculate the equilibrium state with
specified pressure $p_\spec$ and entropy $s_\spec$. To determine the
temperature and the phase fractions, the
\texorpdfstring{$p{s}$}{ps}-flash must be solved.

\citet{Giljarhus2012} considered these equilibrium problems for
single-phase gas and liquid, and two-phase gas-liquid. The extension of the
solution procedures to account for dry-ice along the sublimation line is
described thoroughly by \citet{Hammer2013}, for both the
\texorpdfstring{${e}{v}$}{ev}-flash and the
\texorpdfstring{$p{s}$}{ps}-flash. The
\texorpdfstring{${s}{v}$}{sv}-flash has not been considered
earlier and will be treated in more detail below. We do not rely on
tabulated values in the numerical procedures, but solve the EOS directly.

\subsubsection{The \texorpdfstring{${s}{v}$}{sv}-flash}
When computing the equilibrium fluid states from the set of variables
$\vec{R}$ available when reconstructing in the local characteristic
variables (see \ref{sec:chars}), we must perform an equilibrium
calculation with specified entropy $s_\spec$ and specific volume ${v}_\spec$. The whole
procedure is analogous to that with specified $e_\spec$ and
$v_\spec$.

In principle, the EOSs we consider here can be expressed in terms of the
specific Helmholtz free energy as a function of temperature and density,
$a(T,\rho)$. All other thermodynamic properties can be written in terms of
$a$ and its derivatives. Thus, in the gas-liquid case, solving the equation
set
\begin{equation}
  \label{eq:saturation_pressure}
  \vec{\varphi} \left(\rho_\gas, \rho_\liq \right) =
  \begin{bmatrix}
    \mu \left(T, \rho_\gas \right)
      - \mu \left( T, \rho_\liq \right) \\
      p\left( T, \rho_\gas \right) - p\left( T, \rho_\liq \right)
  \end{bmatrix}
  = \vec{0},
\end{equation}
yields $\rho_\gas^\sat$ and $\rho_\liq^\sat$ for a given $T$. $\mu$ denotes
the chemical potential. Here and in the following, $\vec{\varphi}$ is a
general set of thermodynamic relations which form the left-hand side of an
equation set to be solved.
With the phase densities as functions of temperature, the entropy and
specific volume constraints can be solved for in an outer loop,
\begin{equation}
  \label{eq:glsvflash}
  \vec{\varphi} \left( T, \beta_\gas \right) =
\begin{bmatrix}
  {v}\left( T,  \beta_\gas \right)
    - {v}_\spec \\
  {s}\left( T,  \beta_\gas \right) - {s}_\spec
\end{bmatrix}
= \vec{0},
\end{equation}
to get the equilibrium temperature and the
gas mass fraction $\beta_\gas$. It is also possible to solve
\eqref{eq:saturation_pressure} and \eqref{eq:glsvflash}
simultaneously, but solving them in an nested-loop approach improves
the robustness. The gas-solid equilibrium case is solved in a similar
manner as the gas-liquid case.

In the case where the equilibrium state is single-phase, the EOS
provides the relation $p=p\left(T,1/{v}_\spec\right)$ and we solve
\begin{equation}
  \label{eq:ps_single_phase}
  \varphi \left( T \right) = \left( {s} \left( T \right)
    - {s}_\spec \right) = 0,
\end{equation}
to obtain the equilibrium temperature.

In the gas-liquid-solid equilibrium case, i.e. the triple point, we solve
\begin{equation}
  \label{eq:glssvflash}
    \vec{\varphi} \left( \beta_\gas, \beta_\liq, \beta_\solid \right) =
\begin{bmatrix}
  {v}^\tr_\gas \beta_\gas + {v}^\tr_\liq \beta_\liq + {v}^\tr_\solid \beta_\solid
    - {v}_\spec \\
  {s}^\tr_\gas \beta_\gas + {s}^\tr_\liq \beta_\liq + {s}^\tr_\solid \beta_\solid
  - {s}_\spec \\
  \beta_\gas + \beta_\liq + \beta_\solid - 1
\end{bmatrix}
= \vec{0},
\end{equation}
to obtain the phase mass fractions $\beta_\gas$, $\beta_\liq$ and
$\beta_\solid$. Properties with superscript \tr~are evaluated for the
triple-point pressure and temperature.

\subsection{Speed of sound}
The speed of sound $c$ for a single-phase fluid is computed as
\begin{equation}
  c = \sqrt{\pdc{p}{\rho}{s}}.
\end{equation}
For a gas-liquid mixture in equilibrium, the mixture speed of sound can be
calculated from the combined ${s}{v}$-flash condition
\eqref{eq:glsvflash}, and the modified saturation line condition
\eqref{eq:saturation_pressure}, resulting in the following system of equations:
\begin{equation}
  \label{eq:saturation}
  \vec{\varphi} \left( T, p, \beta_\gas, \rho_\gas, \rho_\liq \right) =
  \begin{bmatrix}
    \mu \left( T, \rho_\gas \right)
      - \mu \left( T, \rho_\liq \right) \\
      p - p\left( T, \rho_\gas \right) \\
      p - p\left( T, \rho_\liq \right) \\
      {v}\left( T, \beta_\gas, \rho_\gas, \rho_\liq \right)
      - {v}_\spec \\
      {s}\left( T, \beta_\gas, \rho_\gas, \rho_\liq \right)
      - {s}_\spec
  \end{bmatrix}
  = \vec{0}.
\end{equation}
The solution $\vec{\chi} = \left[ T, p, \beta_\gas, \rho_\gas, \rho_\liq
\right]^\transp$ to \eqref{eq:saturation} gives a relation for
${v}_\spec$ and ${s}_\spec$:
\begin{equation}
  \vec{\varphi} \left( \ \vec{\chi} \left(
      {v}_\spec, {s}_\spec \right), {v}_\spec,
    {s}_\spec \right) = \vec{0}.
\end{equation}
Differentiating with respect to ${v}_\spec$, we obtain
\begin{equation}
  \label{eq:sos}
  \partial_{\vec{\chi}} \vec{\varphi} \ \partial_{{v}_\spec} \vec{\chi} +
  \partial_{{v}_\spec} \vec{\varphi} = \vec{0},
\end{equation}
whose solution $\partial_{{v}_\spec} \vec{\chi}$
gives an isentropic $\partial_{{v}_\spec} p$, that
can readily be used to calculate the mixture speed of sound. A similar
approach is used to calculate the speed of sound for a gas-solid
mixture. For coexistence of solid, gas and liquid, the equilibrium
model predicts that the speed of sound is zero, since the density can
change isentropically without a change in pressure
\citep[Sec.~2.8.1]{Henderson2000}. Hence, at the triple point, the HEM
loses hyperbolicity. Although this behaviour is believed to be
unphysical, it has not caused practical problems in the present simulations.

\subsection{Parallelization}
It is a computationally intensive task to solve any CFD problem in three
dimensions. Advanced thermodynamic models, like the EOSs used
here, add even more to the computational load. It is therefore necessary to run
simulations in parallel on high-performance computing (HPC) machines.

Since we here use explicit time integration methods, the parallelization becomes
relatively straightforward. In particular, we apply a blockwise domain
decomposition~\cite{Toselli05} to the spatial domain, and the spatial
discretization is applied to the subdomains which are distributed over the nodes
of the HPC cluster using MPI for communication. Due to the width of the WENO
stencils, we require three ghost cells both on the physical domain boundaries
and on the internal boundaries of each subdomain. The values in the ghost cells
are synchronized as necessary, e.g. before each substep in the temporal
discretization.

The implementation of the domain decomposition is based on
PETSc~\citep{Petsc-web-page,Petsc-efficient}. In particular, we follow the
minimally-intrusive parallelization strategy of
\citet{Ervik2014}\footnote{See also the example code dm/ex13f90 included with
PETSc.}, where awareness of the decomposed nature of the domain is hidden from 
the majority of the code. 

\section{Validation}
\label{sec:valid}

\subsection{Advection-diffusion}
To demonstrate the high-order convergence of the numerical methods for
diffusive fluxes, we first consider a smooth problem with the 2D
constant-coefficient advection-diffusion equation,
\begin{equation}
  \label{eq:advec-diff}
  \partial_t q + u_x \partial_x q  + u_y \partial_y q
  = D \left( \partial_x \partial_x q + \partial_y \partial_y q \right).
\end{equation}
Here $q$ is some quantity subject to advective and diffusive
transport, $u_x$ and $u_y$ are the advection velocities and $D$ is the
diffusion coefficient. If we take the entire $x$-$y$ plane as our domain,
an analytical solution to this equation is the Gaussian pulse,
\begin{equation}
  \label{eq:advec-diff_analytical}
  q \left(x, y, t \right) =  \frac{1}{2 \pi \sigma^2(t)}
  \exp \left(-\frac{(x - u_xt - x_0)^2}{2 \sigma^2 (t)}
    -\frac{(y - u_yt - y_0)^2}{2 \sigma^2 (t)} \right).
\end{equation}
Herein, $x_0$ and $y_0$ define the initial position of the Gaussian
pulse and the spread of the pulse $\sigma^2(t)$ is a function of time
$t$ and initial the spread $\sigma^2_0$ at $t = \SI{0}{\second}$,
\begin{equation}
  \sigma^2(t) = 2Dt + \sigma^2_0.
\end{equation}

We consider the specific problem where $x_0 = y_0 = \SI{3/8}{\meter}$,
$\sigma_0 = \SI{1/16}{\meter}$, $u_x = u_y =
\SI{1}{\meter\per\second}$, and $D =
\SI{5e-3}{\meter\squared\per\second}$. We solve the problem inside a
periodic domain $[0,1]\times[0,1]\,\si{\meter}$ and note that, with
the specified parameters, the value of the solution at the domain
boundaries will be much smaller than any numerical errors. The initial
condition follows from \eqref{eq:advec-diff_analytical} evaluated at
$t = \SI 0 \second$.

The governing equation \eqref{eq:advec-diff} is integrated to $t
= \SI{0.25}{\second}$. To eliminate the effect of the RK method on the
convergence order, we use a constant time step of $\SI{e-3}{\second}$ for all
grid sizes. This corresponds to a CFL number of 0.2 on the $200\times200$
grid.

\begin{figure}[tpb]
  \centering
  \includegraphics[width=0.45\textwidth]{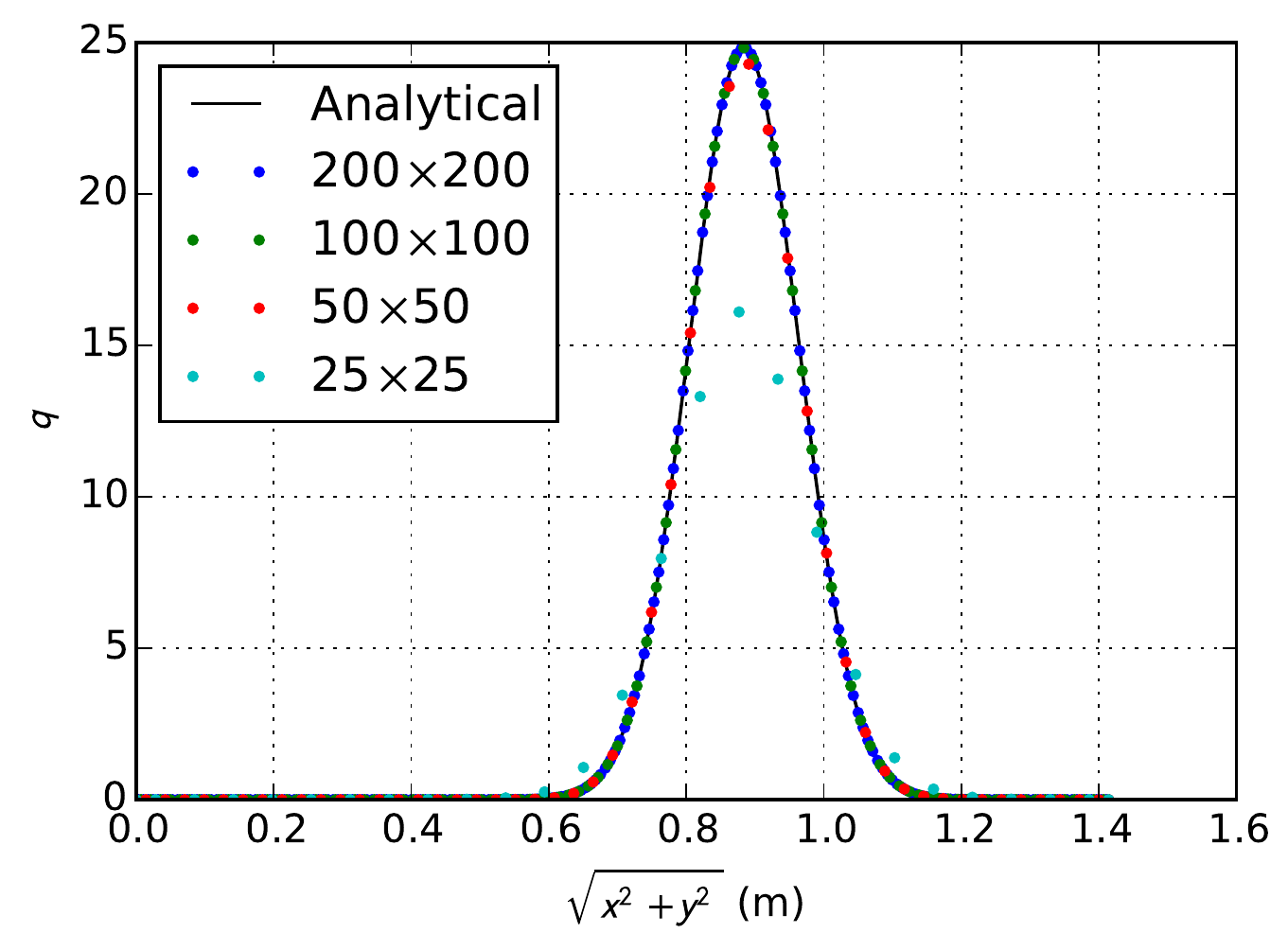}
  \caption{Comparison of the analytical solution and numerical results for the
    advection-diffusion problem. Values along the line $x = y$ are plotted.}
  \label{fig:advec-diffusion}
\end{figure}

The analytical and numerical results at $t = \SI{0.25}{\second}$ are
plotted in Figure \ref{fig:advec-diffusion}. It is evident that the
numerical solutions converge rapidly to the analytical solution when the
grid is refined. The errors and the estimated convergence orders are
presented in Table \ref{tbl:advec-diff}. The errors computed with the
$L_1$-norm are normalized with respect to the number of grid cells. These
results show fifth-order accuracy of the numerical method, also when
treating both diffusive and advective fluxes. The convergence order is
better than what we should expect, since we use a fourth-order accurate
quadrature rule in integrating the fluxes.

The results presented in this section extend the results shown by
\citet{Coralic2014}. In their paper, convergence orders are only
calculated, and shown to be of fifth-order, for the isentropic vortex
case in the absence of any diffusive fluxes. By applying the numerical
schemes to the advection-diffusion equation, we have demonstrated
fifth-order convergence also when diffusive fluxes are included.

\begin{table}[tpb]
  \centering\footnotesize
  \caption{Errors and estimated convergence orders in the
    2D advection-diffusion case.}
  \label{tbl:advec-diff}
  \begin{tabular}{l l l l l}
    \toprule
    Grid & $L_1$-error & $L_1$-order & $L_\infty$-error & $L_\infty$-order \\
    \midrule
    $25\times25$   & \SI{3.208e-01}{} & -
                   & \SI{8.637e+00}{} & - \\
    $50\times50$   & \SI{1.112e-02}{} & \SI{4.85}{}
                   & \SI{4.536e-01}{} & \SI{4.25}{} \\
    $100\times100$ & \SI{2.543e-04}{} & \SI{5.45}{}
                   & \SI{1.060e-02}{} & \SI{5.42}{} \\
    $200\times200$ & \SI{8.953e-06}{} & \SI{4.83}{}
                   & \SI{3.674e-04}{} & \SI{4.85}{} \\
    \bottomrule
  \end{tabular}
\end{table}

\subsection{Isentropic vortex}
To validate the convergence order of the numerical schemes when
applied to the fluid model with a realistic EOS, we next consider a
smooth, inviscid test problem. In particular, we consider a
generalization of the isentropic vortex where the initial condition of
the problem can be found for a general EOS.  The isentropic vortex
with ideal gas was studied by \citet{Balsara2000}, by
\citet{Titarev2004} and by \citet{Coralic2014} to demonstrate the
convergence order of their numerical schemes. In the following
simulations, we first use the EOS of \citet{Peng1976} for pure \coto. Next,
we consider the \citet{Span1996} EOS.

To define the initial condition, we demand a uniform entropy $s$ and prescribe a
rotating velocity field,
\begin{align}
  u_x &= u_{x,\infty} - \frac{\epsilon y}{2 \pi r_0}
      \exp\left(\frac{ 1 - (r/r_0)^2 }{2}\right), \\
  u_y &= u_{y,\infty} + \frac{\epsilon x}{2\pi r_0}
      \exp\left(\frac{ 1 - (r/r_0)^2 }{2}\right).
\end{align}
Herein $u_{x,\infty}$ and $u_{y,\infty}$ are constant background velocities,
$\epsilon$ is the vortex strength and $r_0$ is the vortex radius. We let
$u_{x,\infty} = u_{y,\infty} = \SI{0}{\meter\per\second}$,
$\epsilon = \SI{1000}{\meter\per\second}$ and $r_0 = \SI{20}{\meter}$.

Further, we demand that the pressure gradient give a
centripetal force, whose magnitude and direction keep each fluid
element moving in a circular orbit around the centre of the
vortex. This results in a low-pressure region in the centre. For the
ideal-gas EOS, the required pressure can be determined explicitly
as a function of the radius from centre $r$. For a general EOS,
however, we must numerically integrate the ODE
\begin{equation}
  \label{eq:vortex_ode}
  \od{p}{r} = \frac{\rho \epsilon^2 r}{4 \pi^2 r_0^2}
  \exp \left( 1 - (r/r_0)^2 \right),
\end{equation}
from $r = \infty$ to $r = 0$, with initial condition
$p(r=\infty)=p_\infty$, to obtain the pressure profile. The density
$\rho$ is found in each step of the integration from a $ps$-flash, see Section~\ref{sec:flashes}.

We consider two different cases,
\begin{itemize}
  \item[(i)] a single-phase case where the fluid is in a gaseous state
    everywhere, and
  \item[(ii)] a two-phase case where the fluid is in a gaseous state at $r
    = \infty$, but condenses and enters a gas-liquid state near the centre of
    the vortex.
\end{itemize}

\begin{figure}[t]
  \centering
  \includegraphics[width=0.45\textwidth]{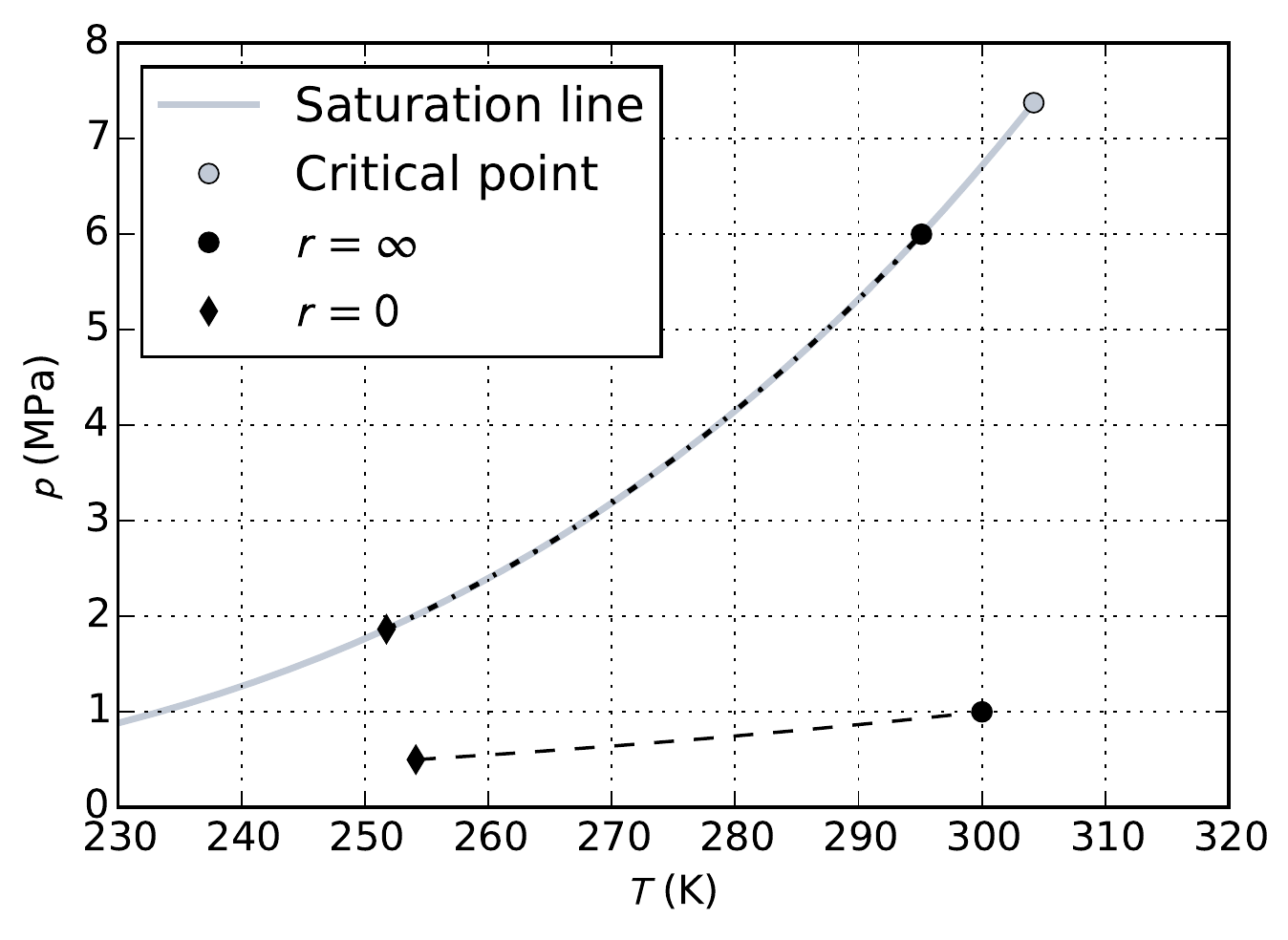}
  \caption{The pressure profiles of the isentropic vortex case,
    plotted in relation to the saturation pressure of \coto. The
    pressure in the single-phase case is drawn as a dashed line and
    the pressure in the two-phase case is drawn as a dash-dotted
    line.}
  \label{fig:vortex_diagram}
\end{figure}

\begin{figure}[t]
  \centering
  \includegraphics[width=0.5\textwidth]{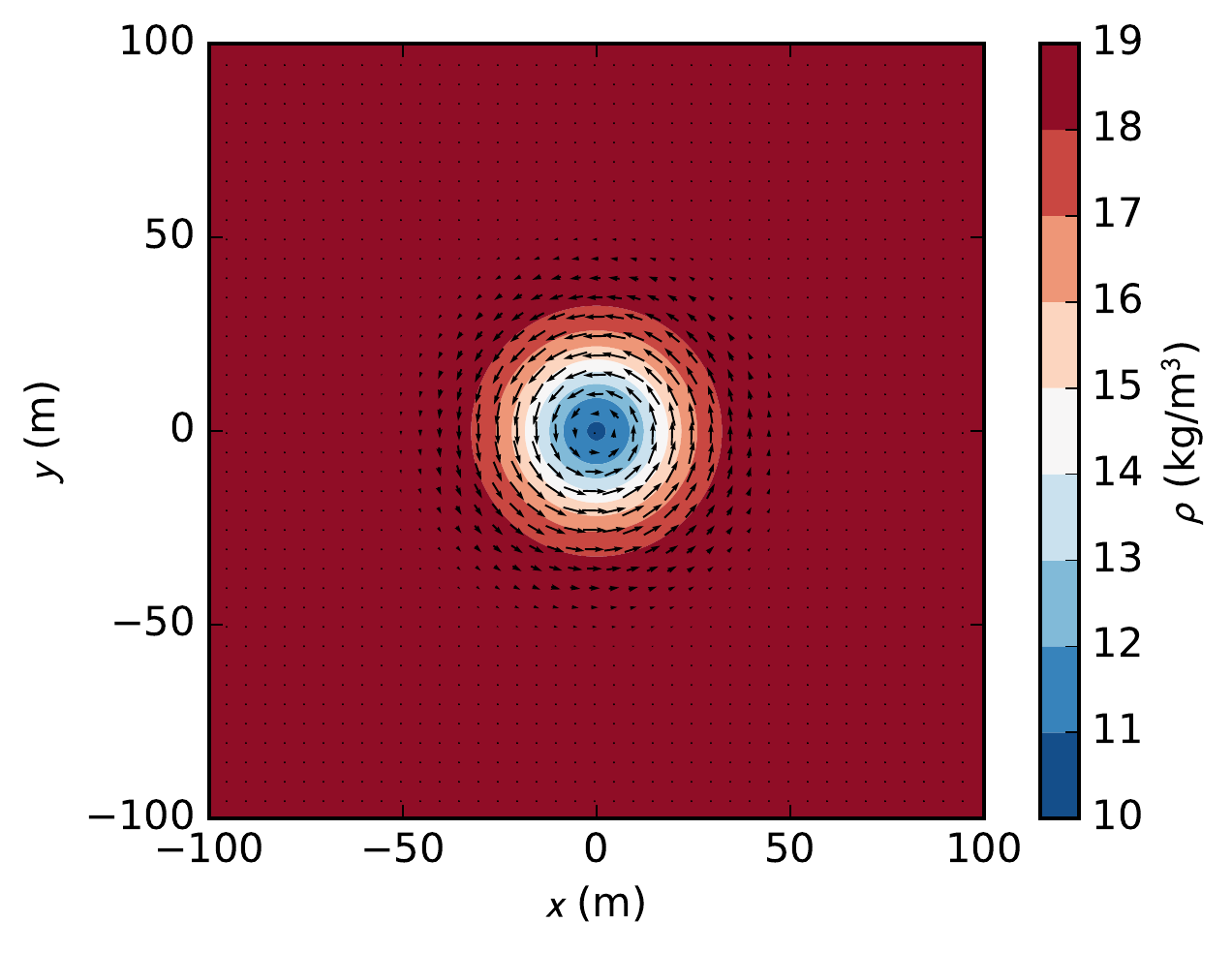}
  \caption{Density contours of the single-phase isentropic vortex case. The
    fluid circulates around the central low-pressure region.}
  \label{fig:vortex_rho_single_phase}
\end{figure}

In the single-phase case, we let $p_\infty =
\SI{1}{\mega\pascal}$. The uniform entropy $s$ is calculated at
$p_\infty$ and $T_\infty = \SI{300}{\kelvin}$. With this reference
state, the pressure profile obtained form \eqref{eq:vortex_ode}
follows an isentrope in the phase diagram that both starts ($r =
\infty$) and ends ($r = 0$) in the gas region (see the dashed line in
Figure \ref{fig:vortex_diagram}). Density contours are plotted Figure
\ref{fig:vortex_rho_single_phase}.

\begin{figure}[t]
  \centering
  \includegraphics[width=0.5\textwidth]{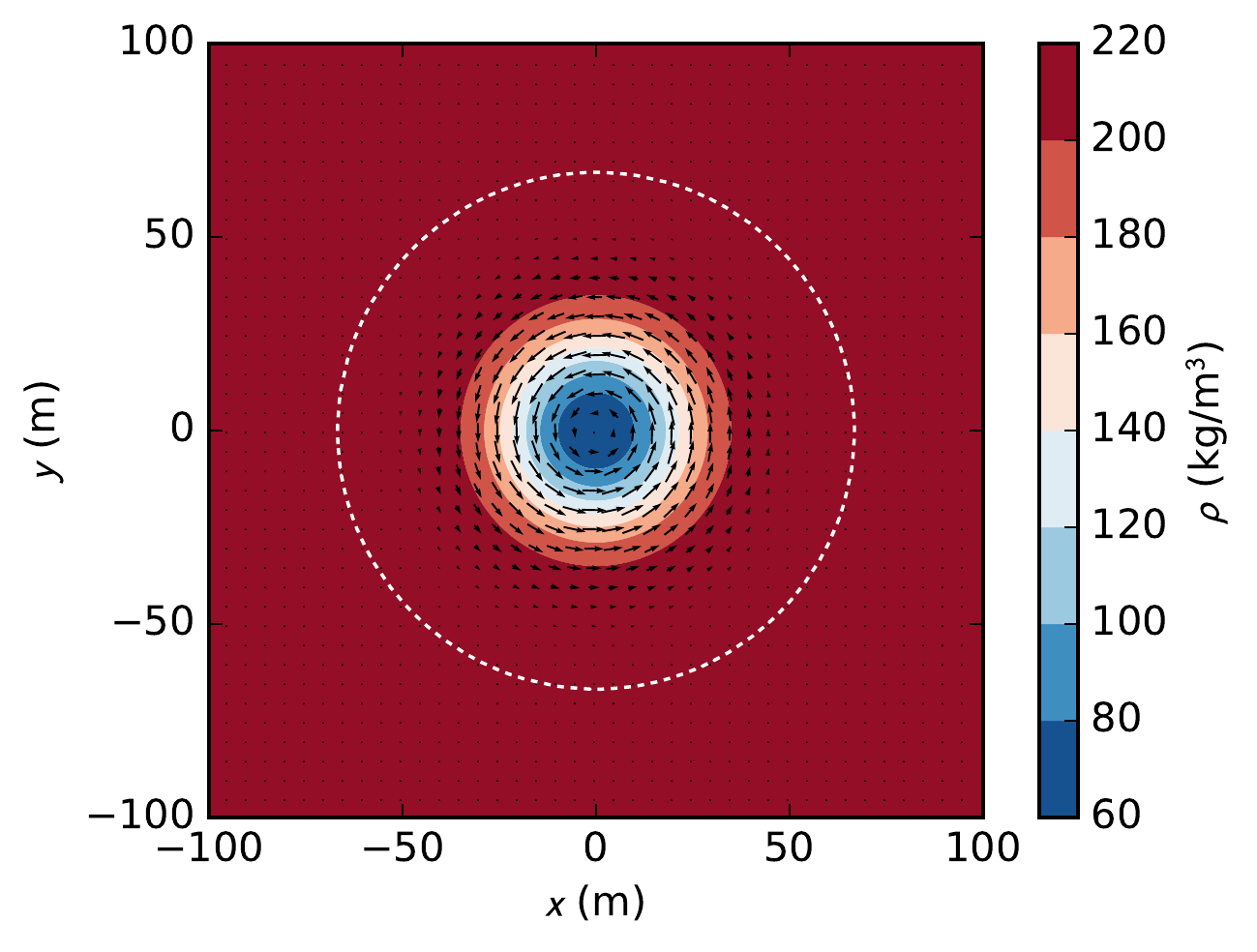}
  \caption{Density contours of the two-phase isentropic vortex
    case. The fluid is in a pure gas state at $r = \infty$, but
    condenses and enters a gas-liquid state near the centre of the
    vortex due to the lower pressure. The white dashed line indicates
    a liquid volume fraction of \SI{e-12}{}.}
  \label{fig:vortex_rho_two_phase}
\end{figure}

In the two-phase case, we let $p_\infty = \SI{6}{\mega\pascal}$. The
entropy $s$ is taken to be the gas entropy at the saturation temperature
corresponding to $p_\infty$. Thus we have $T_\infty =
\SI{295.1}{\kelvin}$. Using this reference state, we get a fluid which is
in a gaseous state at $r = \infty$, but condenses and enters a gas-liquid
state near the centre of the vortex due to the lower pressure. The pressure
profile from \eqref{eq:vortex_ode} follows the saturation line in the phase
diagram (see the dash-dotted line in Figure \ref{fig:vortex_diagram}). The
two-phase region is shown along with density contours in Figure
\ref{fig:vortex_rho_two_phase}.
The solution to the isentropic vortex problem is stationary in the sense
that although we have flow, the values of the state and primitive variables
do not change in time.

It is important to note that in setting the state variables in each cell
initially, we must calculate the state variables at every quadrature point
in the cell and then take the average. Using only one quadrature point per cell
results in a second-order error in the initial condition.
The simulations were run to $t = \SI{0.1}{\second}$ in a square domain
$[-100, 100] \times [-100, 100]\,\si{\meter}$ with periodic boundary
conditions and with a CFL number $C_\cfl = 0.45$. For comparison, reconstruction in
both the set $\vec{W} = \left[e,\rho,u_x,u_y\right]$ and in the
local characteristic variables (see \ref{sec:chars}) was
performed.

\begin{table}[tpb]
  \centering\footnotesize
  \caption{Errors in the density field and estimated convergence orders in the
    single-phase isentropic vortex case (i) with the PR EOS and
    reconstruction in local characteristic variables.}
  \label{tbl:isentropic_vortex_chars_single_phase}
  \begin{tabular}{l l l l l}
    \toprule
    Grid & $L_1$-error & $L_1$-order & $L_\infty$-error & $L_\infty$-order  \\
    \midrule
    $25\times25$   & \SI{1.286e-02}{} & -
                   & \SI{1.432e-01}{} & - \\
    $50\times50$   & \SI{5.975e-04}{} & \SI{4.43}{}
                   & \SI{1.008e-02}{} & \SI{3.83}{} \\
    $100\times100$ & \SI{2.044e-05}{} & \SI{4.87}{}
                   & \SI{3.760e-04}{} & \SI{4.74}{} \\
    $200\times200$ & \SI{7.410e-07}{} & \SI{4.79}{}
                   & \SI{1.140e-05}{} & \SI{5.04}{} \\
    \bottomrule
  \end{tabular}
\end{table}

\begin{table}[tpb]
  \centering\footnotesize
  \caption{Errors in the density field and estimated convergence orders in the
    single-phase isentropic vortex case (i) with the PR EOS and
    reconstruction in $\vec{W} = [e,\rho,u_x,u_y]$.}
  \label{tbl:isentropic_vortex_erhou_single_phase}
  \begin{tabular}{l l l l l}
    \toprule
    Grid & $L_1$-error & $L_1$-order & $L_\infty$-error & $L_\infty$-order  \\
    \midrule
    $25\times25$   & \SI{1.314e-02}{} & -
                   & \SI{1.532e-01}{} & - \\
    $50\times50$   & \SI{6.468e-04}{} & \SI{4.34}{}
                   & \SI{1.225e-02}{} & \SI{3.64}{} \\
    $100\times100$ & \SI{2.196e-05}{} & \SI{4.88}{}
                   & \SI{3.722e-04}{} & \SI{5.04}{} \\
    $200\times200$ & \SI{8.070e-07}{} & \SI{4.77}{}
                   & \SI{1.254e-05}{} & \SI{4.89}{} \\
    \bottomrule
  \end{tabular}
\end{table}

\begin{table}[btp]
  \centering\footnotesize
  \caption{Errors in the density field and estimated convergence orders in the
    two-phase isentropic vortex case (ii) with the PR EOS and
    reconstruction in the local characteristic variables.}
  \label{tbl:isentropic_vortex_chars_two_phase}
  \begin{tabular}{l l l l l}
    \toprule
    Grid & $L_1$-error & $L_1$-order & $L_\infty$-error & $L_\infty$-order  \\
    \midrule
    $25\times25$   & \SI{1.398e-01}{} & -
                   & \SI{1.775e+00}{} & - \\
    $50\times50$   & \SI{8.523e-03}{} & \SI{4.04}{}
                   & \SI{1.365e-01}{} & \SI{3.70}{} \\
    $100\times100$ & \SI{3.048e-04}{} & \SI{4.81}{}
                   & \SI{5.926e-03}{} & \SI{4.53}{} \\
    $200\times200$ & \SI{1.119e-05}{} & \SI{4.77}{}
                   & \SI{3.183e-04}{} & \SI{4.22}{} \\
    \bottomrule
  \end{tabular}
\end{table}

\begin{table}[btp]
  \centering\footnotesize
  \caption{Errors in the density field and estimated convergence orders in the
    two-phase isentropic vortex case (ii) with the PR EOS and
    reconstruction in $\vec{W} = [e,\rho,u_x,u_y]$.}
  \label{tbl:isentropic_vortex_erhou_two_phase}
  \begin{tabular}{l l l l l}
    \toprule
    Grid & $L_1$-error & $L_1$-order & $L_\infty$-error & $L_\infty$-order  \\
    \midrule
    $25\times25$   & \SI{1.615e-01}{} & -
                   & \SI{1.976e+00}{} & - \\
    $50\times50$   & \SI{9.200e-03}{} & \SI{4.13}{}
                   & \SI{1.591e-01}{} & \SI{3.63}{} \\
    $100\times100$ & \SI{3.314e-04}{} & \SI{4.80}{}
                   & \SI{6.635e-03}{} & \SI{4.58}{} \\
    $200\times200$ & \SI{1.215e-05}{} & \SI{4.77}{}
                   & \SI{2.522e-04}{} & \SI{4.72}{} \\
    \bottomrule
  \end{tabular}
\end{table}

The errors in the density field and the estimated convergence orders
for the single-phase case (i) are presented in Table
\ref{tbl:isentropic_vortex_chars_single_phase} for reconstruction in
the local characteristic variables and in Table
\ref{tbl:isentropic_vortex_erhou_single_phase} for reconstruction in
$\vec{W}$.
It is observed that reconstruction in the characteristic variables,
although much more computationally expensive, produces an error that
is of the same order of magnitude as reconstruction in $\vec{W}$. The
general trend is that the errors obtained with characteristic
reconstruction are slightly lower. The difference is small, however,
and only about 8\% in the $L_1$-norm on the $200\times200$
grid. High-order convergence was obtained with both reconstruction
options and significant oscillations were not observed in any
simulations of this case.

Despite using a third-order RK method and a fourth-order quadrature
rule, we get close to fifth-order convergence with both reconstruction
alternatives. Similar behaviour was also observed by
\citet{Titarev2004} and by \citet{Coralic2014} and suggests that the
error in the time integration method is small compared to that of the
spatial discretization in this case.

We have also considered a three-dimensional, single-phase isentropic
vortex problem with rotation in the $x$-$z$ plane. These results also
showed close to fifth-order convergence and are omitted here.

The errors in the density field and the estimated convergence orders
for the two-phase case (ii) are presented in
Table~\ref{tbl:isentropic_vortex_chars_two_phase} for reconstruction
in the local characteristic variables and in
Table~\ref{tbl:isentropic_vortex_erhou_two_phase} for reconstruction
in $\vec{W}$. Again, the general trend is that the errors are slightly
smaller for reconstruction in the local characteristic variables than
for reconstruction in $\vec{W}$, but again the differences are
small. For this case, the difference is about \SI{0.5}{\percent} in
the $L_1$-norm on the $200\times200$ grid, while the error in the
$L_\infty$-norm on the same grid is smaller with reconstruction in
$\vec{W}$. As for the single-phase case, no significant oscillations
were observed and the errors show close to fifth-order covergence.

We have also performed simulations of both case (i) and case (ii) using the
more complex SW EOS, in place of the PR EOS, and reconstruction in
$\vec{W}$. The errors and convergence orders are shown in
Table~\ref{tbl:isentropic_vortex_erhou_single_phase_sw} and Table
\ref{tbl:isentropic_vortex_erhou_two_phase_sw} respectively. For both
cases, the results are similar to those obtained with the PR EOS (see
Table~\ref{tbl:isentropic_vortex_erhou_single_phase} and Table
\ref{tbl:isentropic_vortex_erhou_two_phase}). This indicates that the order
of the numerical method is not affected by the degree of complexity of the
EOS underlying the phase equilibrium calculations.

\begin{table}[btp]
  \centering\footnotesize
  \caption{Errors in the density field and estimated
    convergence orders in the single-phase isentropic vortex case (i)
    with the SW EOS and reconstruction in $\vec{W} = [e,\rho,u_x,u_y]$.}
  \label{tbl:isentropic_vortex_erhou_single_phase_sw}
  \begin{tabular}{l l l l l}
    \toprule
    Grid & $L_1$-error & $L_1$-order & $L_\infty$-error & $L_\infty$-order  \\
    \midrule
    $25\times25$   & \SI{1.303e-2}{} & -
                   & \SI{1.518e-1}{} & - \\
    $50\times50$   & \SI{6.391e-4}{} & \SI{4.35}{}
                   & \SI{1.207e-2}{} & \SI{3.65}{} \\
    $100\times100$ & \SI{2.175e-5}{} & \SI{4.88}{}
                   & \SI{3.663e-4}{} & \SI{5.04}{} \\
    $200\times200$ & \SI{8.001e-7}{} & \SI{4.76}{}
                   & \SI{1.236e-5}{} & \SI{4.89}{}\\
    \bottomrule
  \end{tabular}
\end{table}

\begin{table}[btp]
  \centering\footnotesize
  \caption{Errors in the density field and estimated
    convergence orders in the two-phase isentropic vortex case (ii)
    with the SW EOS and reconstruction in $\vec{W} = [e,\rho,u_x,u_y]$.}
  \label{tbl:isentropic_vortex_erhou_two_phase_sw}
  \begin{tabular}{l l l l l}
    \toprule
    Grid & $L_1$-error & $L_1$-order & $L_\infty$-error & $L_\infty$-order  \\
    \midrule
    $25\times25$   & \SI{2.008e-1}{} & -
                   & \SI{2.402}{} & - \\
    $50\times50$   & \SI{1.150e-2}{} & \SI{4.13}{}
                   & \SI{1.983e-1}{} & \SI{3.60}{}  \\
    $100\times100$ & \SI{4.081e-4}{} & \SI{4.82}{}
                   & \SI{7.891e-3}{} & \SI{4.65}{} \\
    $200\times200$ & \SI{1.467e-5}{} & \SI{4.80}{}
                   & \SI{3.055e-4}{} & \SI{4.69}{} \\
    \bottomrule
  \end{tabular}
\end{table}

To summarize, close to fifth-order convergence is observed in both the
single-phase (i) and the two-phase (ii) isentropic vortex cases. This
demonstrates that the high-order convergence of the numerical methods
is not limted to single-phase problems with simple EOS. The results
also suggest that in this case, errors in third-order temporal
integration and fourth-order quadrature rules are not dominating.
Differences in the error between reconstruction in $\vec{W}$ and local
characteristic variables are small. As reconstruction in $\vec{W}$ is
much less computationally intensive, it may be preferable in cases
where the more advanced option is not needed in order to avoid
oscillations.

\subsection{Double Mach reflection of strong shock}

\begin{figure*}[tpb]
\begin{subfigure}[t]{\linewidth}
  \centering
  \includegraphics[width=\linewidth]{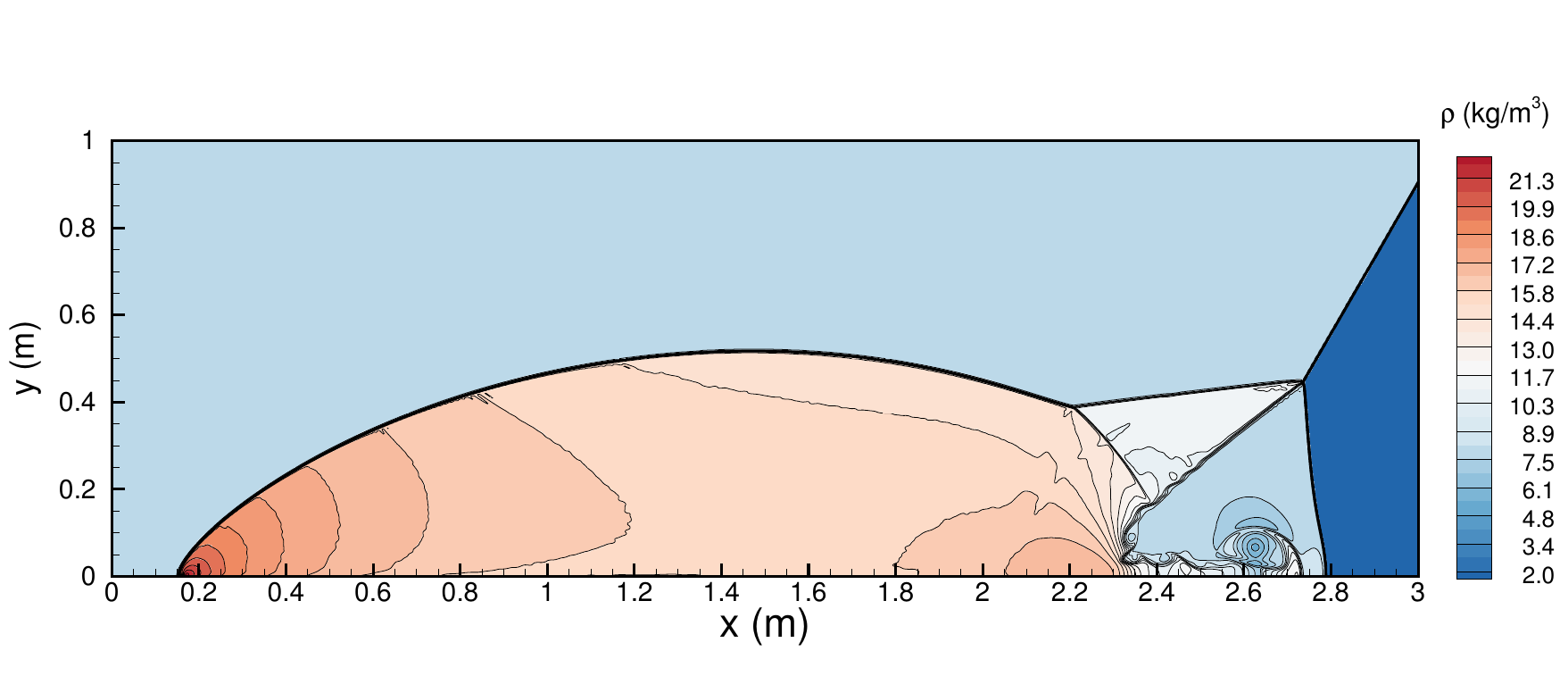}
  \caption{Whole flow field}
\end{subfigure}\\
\begin{subfigure}[t]{\linewidth}
  \centering
  \includegraphics[width=0.45\linewidth]{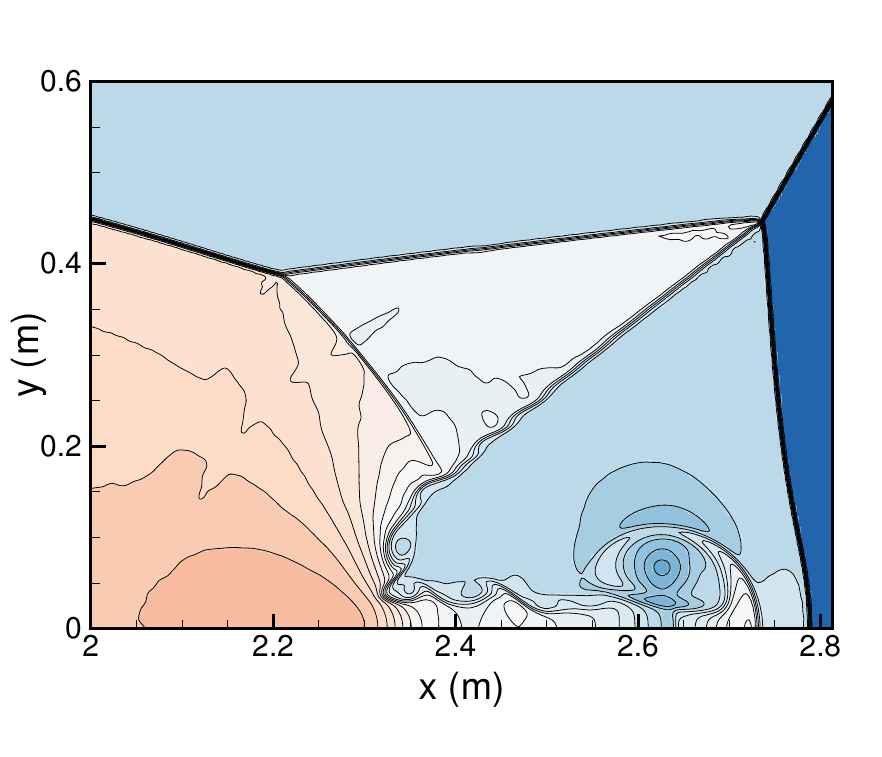}
  \caption{Close-up}
\end{subfigure}
\caption{Contour plot of density in the double Mach reflection
    problem. Thirty contour lines in the range
    $\left[\SI{2}{\kilo\gram\per\meter^3},
      \SI{22}{\kilo\gram\per\meter^3}\right]$ are used.}
  \label{fig:double_mach}
\end{figure*}

Next we consider a double Mach reflection of a strong shock incident on
a planar wall. This problem tests the ability of the numerical
methods, and their implementation, to handle strong shocks. This type
of reflection problems was designed to mimic experiments where a shock
propagates down a tube and hits an inserted wedge. The flow pattern
resulting from the reflection of the shock on the wedge is complicated
and challenging to simulate numerically. In particular, the forward
jet that forms along the wall behind the first Mach stem (from
approximately $x = \SI{2.3}{\metre}$ to $x = \SI{2.8}{\metre}$ in Figure
\ref{fig:double_mach}) is difficult to resolve \cite{Woodward1984}.

To have results that could be compared with those from the literature,
the simulation was run with parameters that seem to be the most
common, e.g.\ the parameters used by \citet{Titarev2004}. This also
implies the ideal-gas EOS and adiabatic constant $\gamma
= 1.4$.

We consider a domain $[0,4] \times [0,1]\,\si{\meter}$. A shock is
initiated with a right-moving Mach 10 front incident on the $x$-axis
at $x=\SI{1/6}{\meter}$ and a forward angle of $60^\circ$. The
undisturbed region in front of the shock is at rest with
$p=\SI{1}{\pascal}$ and $\rho=\SI{1.4}{\kilo\gram\per\meter^3}$. In
post-shock region the fluid moves with velocities
$u_y=\SI{-4.125}{\meter\per\second}$ and $u_x=-\sqrt{3}u_y$, and has
$p=\SI{116.5}{\pascal}$ and $\rho=\SI{8.0}{\kilo\gram\per\meter^3}$.

The south wall is reflecting for $x\geq\SI{1/6}{\meter}$ and has
post-shock values for $x<\SI{1/6}{\meter}$. The eastern boundary has a
zero-gradient boundary condition and the western boundary carries post
shock values. The northern boundary is dynamic with post-shock values
in the post-shock region and undisturbed values in the undisturbed
region. For further details on the problem and how it is defined, the
reader is referred to \citet{Woodward1984}. A note on the setup of the case
is given by \citet{Kemm2016}.

We use a CFL number of $C_\cfl = 0.4$ and a grid size of
$1920\times480$. To avoid spurious oscillations, it was necessary to
perform reconstruction in the local characteristic variables for this case
(see \ref{sec:chars}).

A contour plot of the density at $t = \SI{0.2}{\second}$ is shown in Figure
\ref{fig:double_mach}. We observe that both Mach stems (starting at about
$x=\SI{0.15}{\metre}$ and $x=\SI{2.8}{\metre}$) are sharp and that the
details of the jet structure (in the lower part of the figure around
$x=\SI{2.5}{\metre}$) are well-resolved. The positions of the shocks and
discontinuities compare well with results from the literature
\cite{Woodward1984,Jiang1996,Shi2002,Titarev2004}. The same is true for the
positions of the isodensity lines and the level of detail obtained here,
compared to the fifth-order WENO reconstruction scheme on the same grid
size in \cite{Titarev2004}. We conclude that the solution obtained here is
in good agreement with that which is generally accepted in the literature.

\subsection{Air jet flow from rectangular nozzle}
\label{sec:airjet}

The present validation test consists of a 3D direct numerical simulation of
a transitional subsonic air jet issuing from a rectangular nozzle into
quiescent air. The chosen configuration (with air modelled as an ideal gas)
is computationally less expensive than the main demonstration case of the
present paper, the supersonic \ce{CO2} jet with complex thermodynamic
behaviour (Section~\ref{sec:co2jet}). It retains, however, some of its
general flow features and, due to its straightforward boundary condition
specification, can be conveniently compared with experimental data
presented by \citet{Deo07}.

The air jet issues with a centreline velocity of
$u_{y,\text{c},0} = \SI{26.3}{\meter\per\second}$ into initially
quiescent air at a temperature of \SI{293}{\kelvin} and pressure of
\SI{1}{\atm}. The nozzle configuration consists of a rectangular slot
of dimensions $w \times h$ characterized by high aspect ratio $w/h
\sim 10$ where $h=\SI{5.6}{\milli\meter}$, as in
\cite{Deo07}, resulting in a jet Reynolds number $\Rey_{jet} = \rho u_{y,\text{c},0} h / \eta \sim \num{e5}$.
Furthermore, the imposed jet inlet velocity distribution
follows a ``top-hat'' profile to reproduce the effects of a
conventional smoothly-contracting nozzle shape with laminar boundary
layers, as discussed in \cite{Deo07}. For the experimental conditions
that are targeted in the present work, the jet's own turbulent
velocity fluctuations are considered relatively unimportant and no
velocity perturbations are introduced at the jet inlet. Natural
perturbations of the acoustic field, intrinsically represented by the
present compressible formulation, are sufficient to cause the jet flow
to become unstable and break-up.  The direction of the jet flow is in
the positive $y$-direction. The two-dimensionality of the jet at the
inlet nozzle is ensured by a periodic boundary condition in the
spanwise direction \textit{z} to eliminate border effects. Open,
non-reflecting boundary conditions are imposed at the upper $y$
boundary and at both $x$ boundaries.

\begin{figure}
  \begin{center}
    \adjustbox{trim=1pt,clip}{\includegraphics[width=0.5\linewidth]{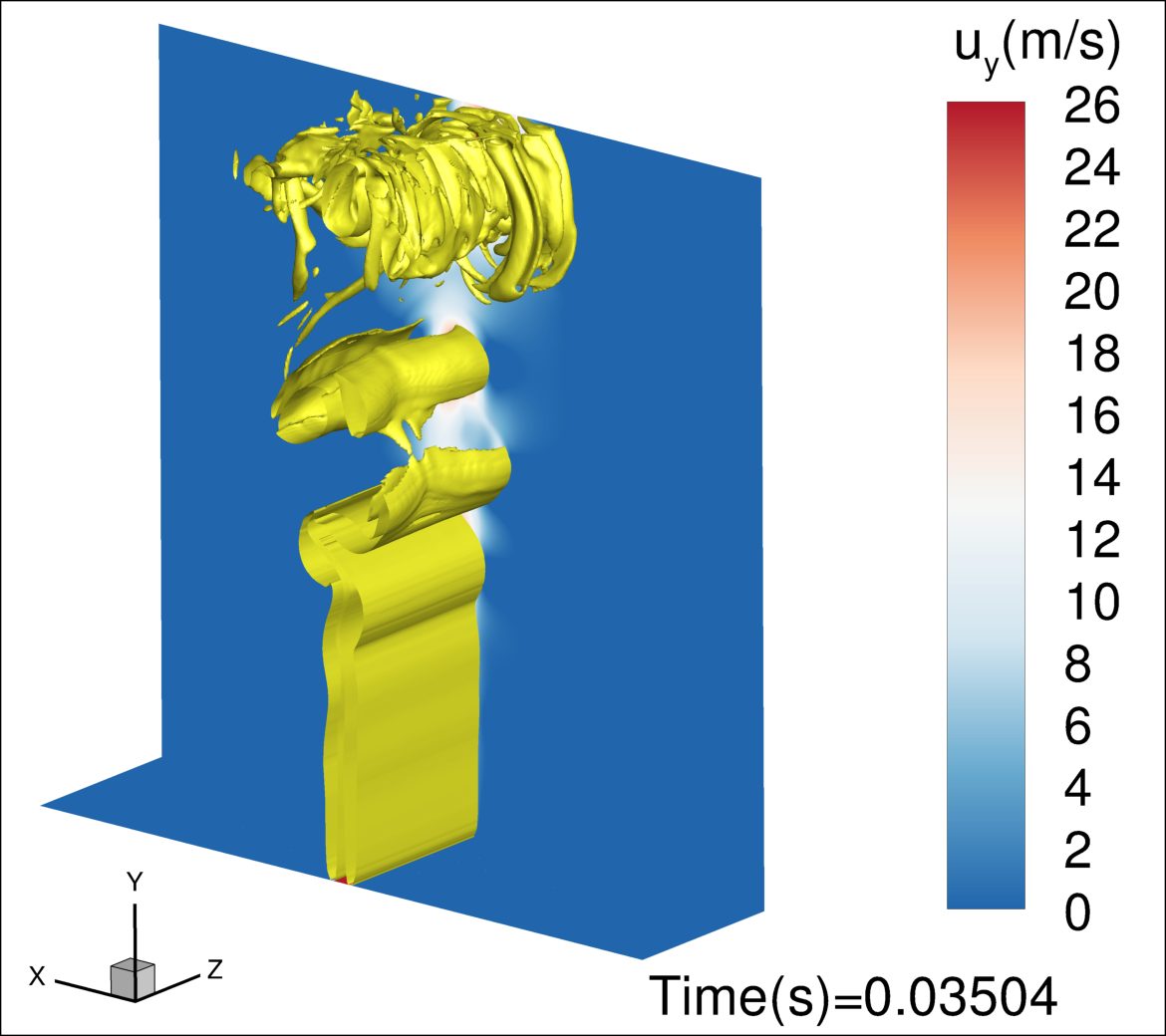}}
  \end{center}
  \caption{Air jet: Visualization of the flow and computational domain. The
    yellow colour indicates the \SI{2500}{\per\second} vorticity magnitude.}
  \label{fig:domain}
\end{figure}

Figure \ref{fig:domain} presents a visualization of the three-dimensional
computational domain and a snapshot of the flow at
$t=\SI{35}{\milli\second}$. The domain extends $L_y=\SI{20}{\centi\meter}$
in the $x$- and $y$-directions and \SI{5}{\centi\meter} in the
$z$-direction. The computational domain is discretized by
$400\times400\times100$ grid nodes in the $x$-, $y$- and $z$-direction,
respectively. This gives a constant spatial resolution of
\SI{0.5}{\milli\meter} throughout the computational domain and implies that
the jet inlet dimension $h$ is resolved by twelve grid nodes, while
$C_\cfl=0.3$ in this three-dimensional simulation.

\begin{figure}
\begin{subfigure}[t]{0.45\linewidth}
  \centering
  \adjustbox{trim=1pt,clip}{\includegraphics[width=\linewidth]{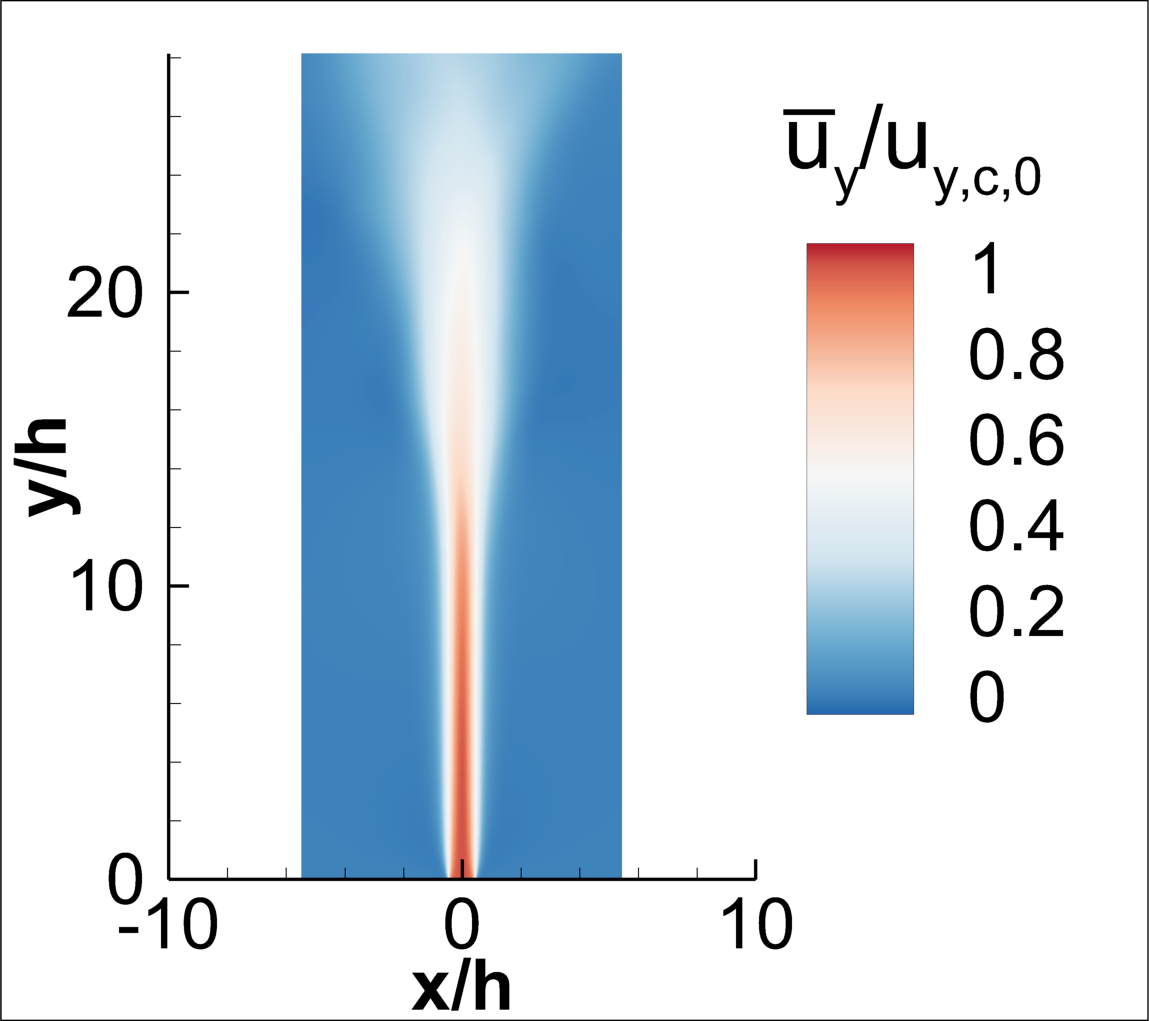}}
  \caption{Average of the velocity component in the jet
    direction, $\bar{u}_y$.}
 \label{fig:2dransvm}
\end{subfigure}
\hfill
\begin{subfigure}[t]{0.45\linewidth}
  \centering
    \adjustbox{trim=1pt,clip}{\includegraphics[width=\linewidth]{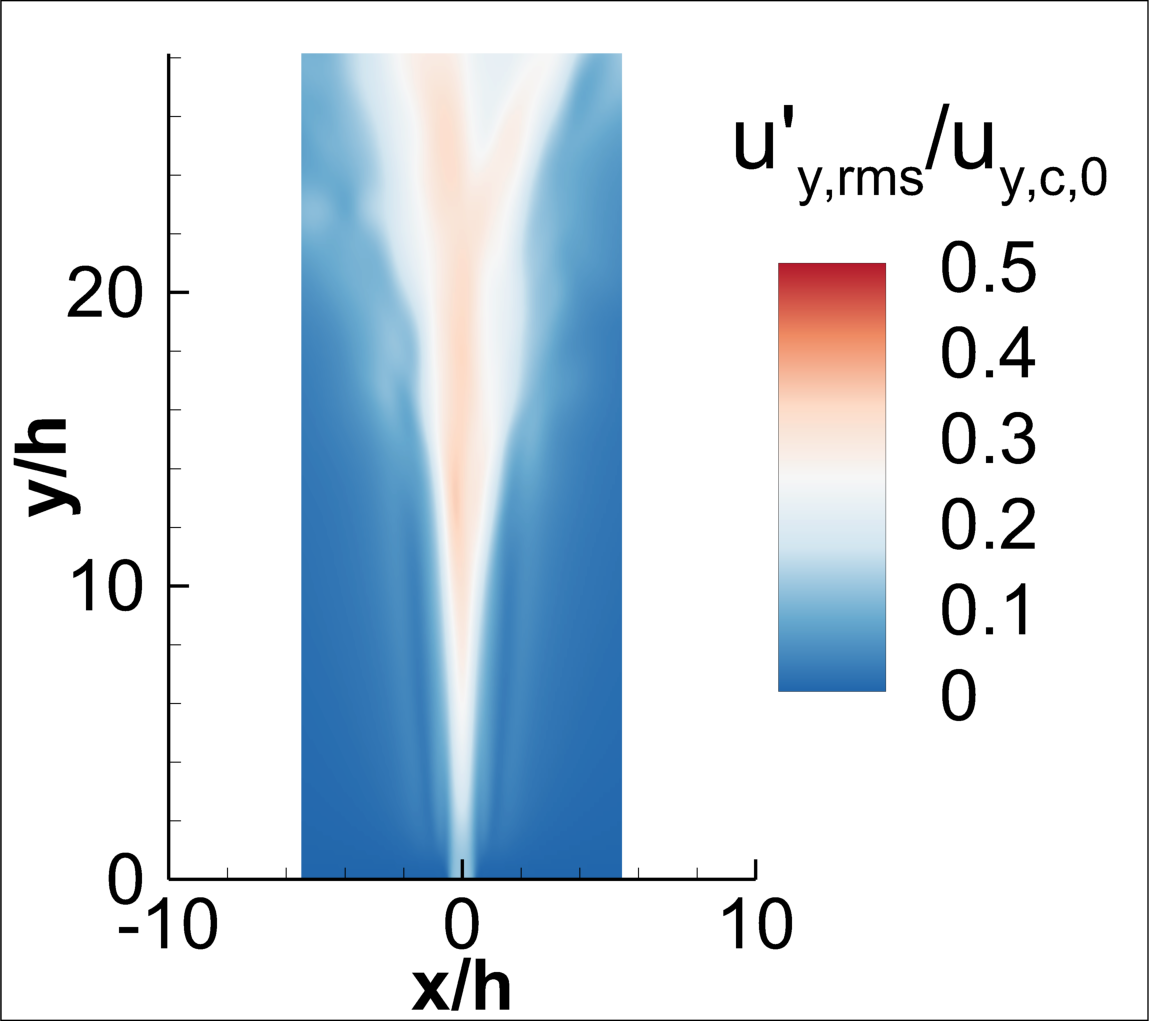}}
  \caption{Average of the root mean square of the velocity fluctuation
    $u'_{y,\text{rms}}$ in the jet direction.}
  \label{fig:2dransvp}
\end{subfigure}
 \caption{Air jet: Pointwise temporal and spatial (in $z$-direction)
    average velocity quantities, normalized by the
    centreline value at the nozzle exit, $u_{y,\text{c},0}$.}
  \label{fig:jetcontours}
\end{figure}

After the simulation is started, a settling time approximately equal to
$t_{\text{ini}} \sim L_y/u_{y,\text{c},0} = \SI{7.6}{\milli\second}$ is
allowed in order to ensure that the initial transient is transported
outside of the computational domain. The time step is allowed to adapt to
the varying CFL conditions. After the initial transient, however, it is
observed that it stabilizes around $\Delta t \sim
\SI{0.4}{\micro\second}$. Sampling of the solution is started for
$t>t_{\text{ini}}$, every 500 time steps or $\SI{0.2}{\milli\second}$
(corresponding approximately to the characteristic jet time $t_{\text{jet}}
= h/u_{y,\text{c},0}$). Sampling is stopped at $t_{\text{end}} =
\SI{50}{\milli\second}$ after approximately 5.5 domain transit times and
238 characteristic jet times $t_{\text{jet}}$. Figure \ref{fig:2dransvm}
illustrates the spatial pattern of the mean jet (wall-normal) velocity
component that is averaged in time and in the homogeneous spanwise
direction ($z$) and normalized by the centreline jet exit velocity
$u_{y,\text{c},0}$. Figure \ref{fig:2dransvp} analogously shows the
normalized root mean square of the velocity fluctuation in the jet
direction. The typical features of jet flows are present with a clearly
visible potential core ($u_y \sim u_{y,\text{c},0}$) characterized by low level of
velocity fluctuations. Downstream of it, as the jet spreads, the velocity
fluctuations increase due to entrainment of the surrounding air and jet
break-up.

\begin{figure}
  \begin{center}
    \includegraphics[width=0.55\linewidth]{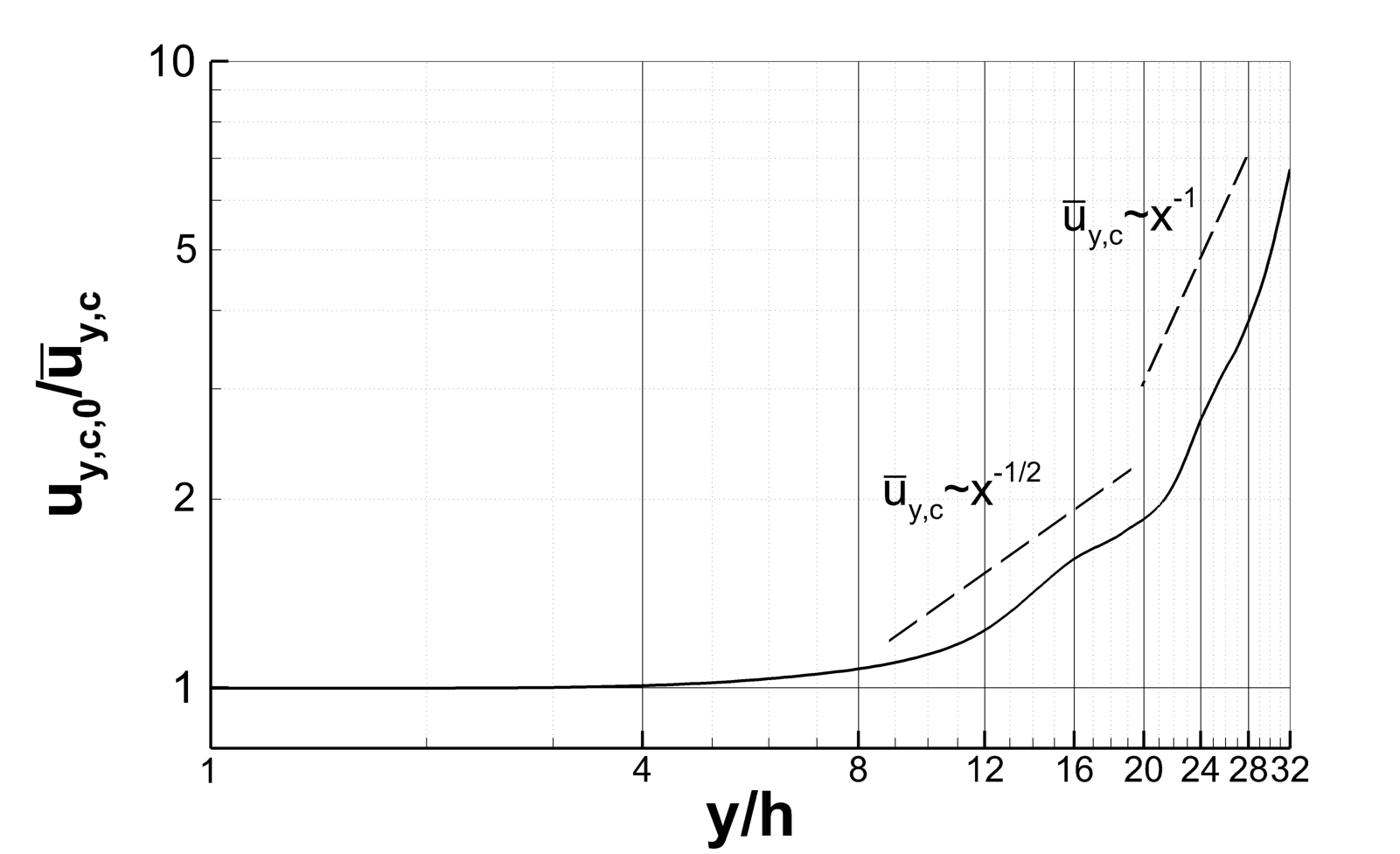}
  \end{center}
  \caption{Air jet: 1D profile of the normalized centreline mean jet
    velocity $u_{y,\text{c},0}/\bar{u}_y$ illustrating the
    well-known decay rates in the jet self-similar region.}
  \label{fig:meanveldecay}
\end{figure}

\begin{figure}
  \begin{center}
    \includegraphics[width=0.55\linewidth]{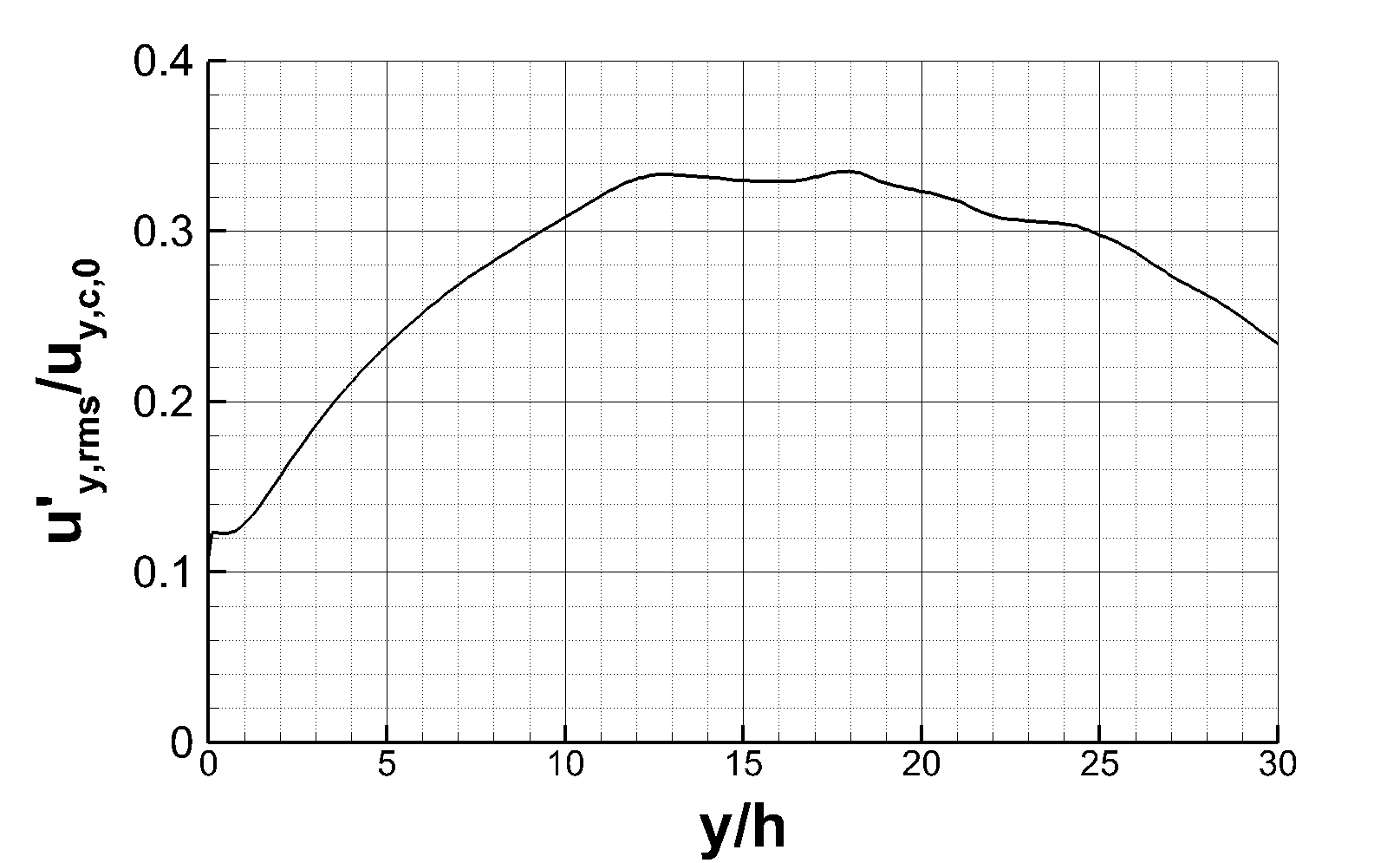}
  \end{center}
  \caption{Air jet: 1D profile of the normalized fluctuating jet velocity
    $u'_{y,\text{rms}}/u_{y,\text{c},0}$ illustrating the well-known
    decay rates in the jet self-similar region and peak in the
    turbulent fluctuations for $y/h \sim 10$.}
  \label{fig:velprimdecay}
\end{figure}

Figure \ref{fig:meanveldecay} provides a more quantitative measure of
the spatial extent of the potential core and of the characteristic
mean velocity decay in the region of self-similarity. The length of
the potential core, equal to $4h$, is in very good accordance with the
values reported by \citet{Deo07}. Furthermore, the
characteristic $1/2$-power law inverse decay can also be clearly
observed, indicating a statistically two-dimensional behaviour of the
jet, together with the appearance, further downstream, of the
inversely-linear decay that indicates transition to a (more
three-dimensional) axisymmetric form \cite{Deo07}. As pointed out in
\cite{Deo07}, the logarithmic scale used in the abscissa is essential
to identify the points of transition from statistically
two-dimensional to three-dimensional mean flow. Although a
considerable spread in the specific spatial location of this
transition is present in the available data, see Figures 6 and 7 in \cite{Deo07},
the present results are consistent with the observed trend that the
transition is delayed for increasing slot aspect ratios and
anticipated for decreasing ones. The present aspect ratio $w/h$ is
approximately 10 with the transition starting 20 nozzle widths $h$
downstream of the jet inlet, while for the aspect ratios of 30 and 60
investigated experimentally by \cite{Deo07}, the same transition starts
at 30 and 50 nozzle widths $h$ downstream of the jet inlet. Finally,
it is worth noticing the very good accordance between the present
results and the results presented in Figure 7 of \cite{Deo07} with
respect to the spatial evolution of the centreline velocity
fluctuations, see Figure \ref{fig:velprimdecay}: the rapid increase
for $y/h$ between 0 and 10, the peak located between 10 and 20 and the
following decay.

\begin{figure}
  \begin{center}
    \includegraphics[width=0.5\linewidth]{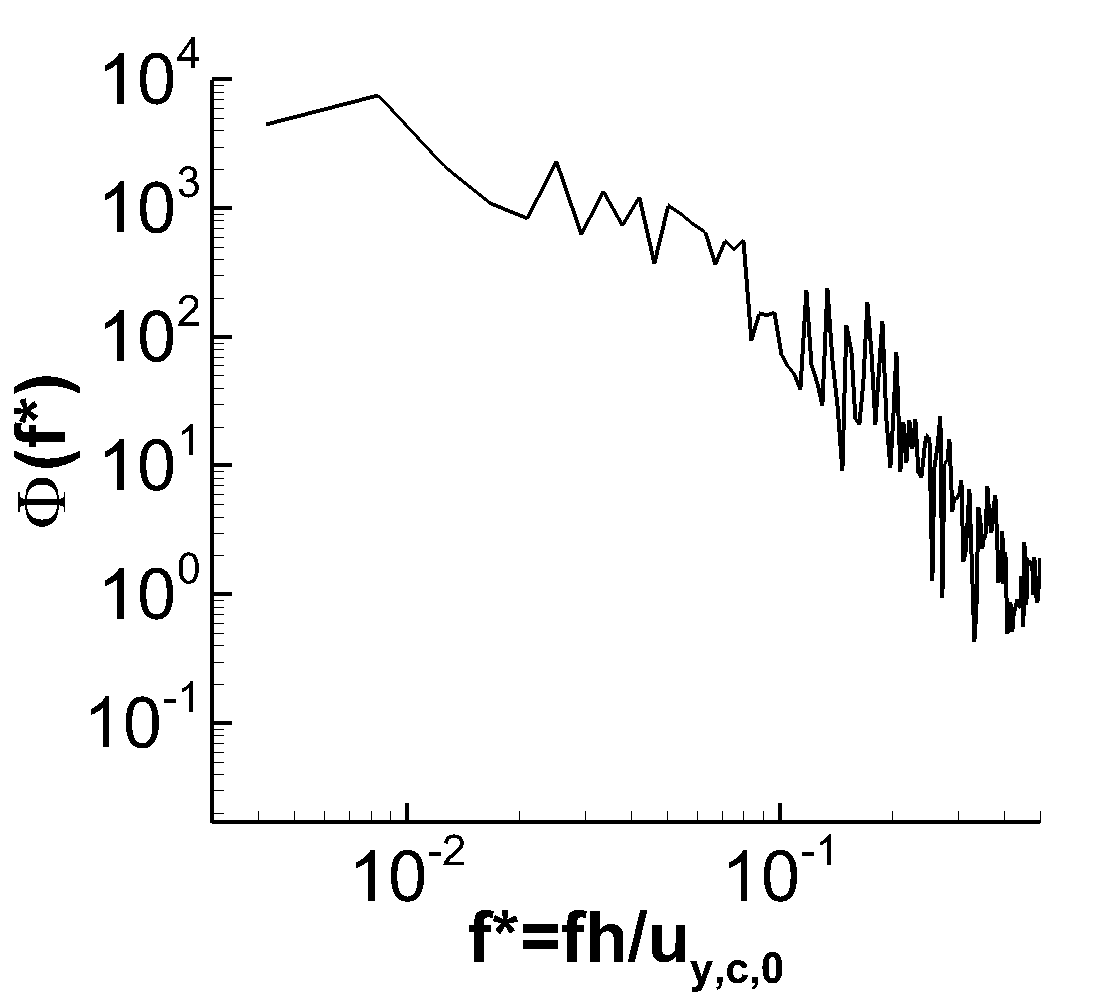}
  \end{center}
  \caption{Air jet: Power spectrum of the instantaneous velocity sampled at
    $y/h=3$ on the centreline of the jet potential core.}
  \label{fig:spectra}
\end{figure}

Figure \ref{fig:spectra} complements the previous analysis by
providing insight into the time-dependent behaviour of the flow
structure in the near field. The power spectrum (distribution of the
energy of a waveform among its different frequency components) of the
fluctuating instantaneous velocity is obtained through a discrete Fourier
transform and plotted versus the frequency, normalized by the characteristic
frequency defined as $1/t_{\text{jet}} = u_{y,\text{c},0}/h$ for a sampling
location within the potential core region (at $y/h=3$).
The presence of broad peaks in the spectrum indicates the generation of regularly-occurring large-scale coherent
vortices, a well-known feature of jet flows. Furthermore, the observed
decay of approximately two orders of magnitude in the spectrum is
consistent with the experimental observations described in~\cite{Deo07}.

In summary, the method seems able to accurately capture the main features of
transitional planar jets at the resolution employed in the present test.

\section{Simulation of \texorpdfstring{\coto}{CO2} jet}
\label{sec:co2jet}
As an application of our method, we now consider two cases involving
high-speed flow of \coto, including phase change and three-phase
(gas-liquid-solid) flow. 

\subsection{Case A}
Some relevant experimental data were presented by
\citet{Li2016}, and we focus on the case labelled ``RHP = 0.030'', see Figure~\ref{fig:co2jet-photo}.

\subsubsection{Case description}
Stagnant pure \coto of temperature $T=\SI{40}{\celsius}$ and pressure
$p=\SI{52.2}{\bar}$ is expanded through a round hole of diameter
$d\ped{e}=\SI{1}{\milli\metre}$. We consider that \coto has replaced air in
the vicinity of the jet, so the initial condition in the computational
domain is stagnant \coto of temperature $T_0=\SI{20}{\celsius}$ and
pressure $p_0=\SI{1}{\bar}$.  At $t=\SI{0}{\second}$, the \coto starts
flowing through the hole.  The jet is highly underexpanded and starts
forming a barrel shock, which attains a steady state after about
$t=\SI{0.25}{\milli\second}$. The Reynolds number at the inlet is \num{6e5}.

In this case, the thermodynamic properties of pure \coto are
calculated as described in Section~\ref{subsec:thermphys}, with the SW
EOS. The thermal conductivity is set to
\SI{0.0145}{\watt\per\metre\per\kelvin}.

\subsubsection{Computational set-up}
\begin{figure}
\centering
\includegraphics[width=0.5\linewidth]{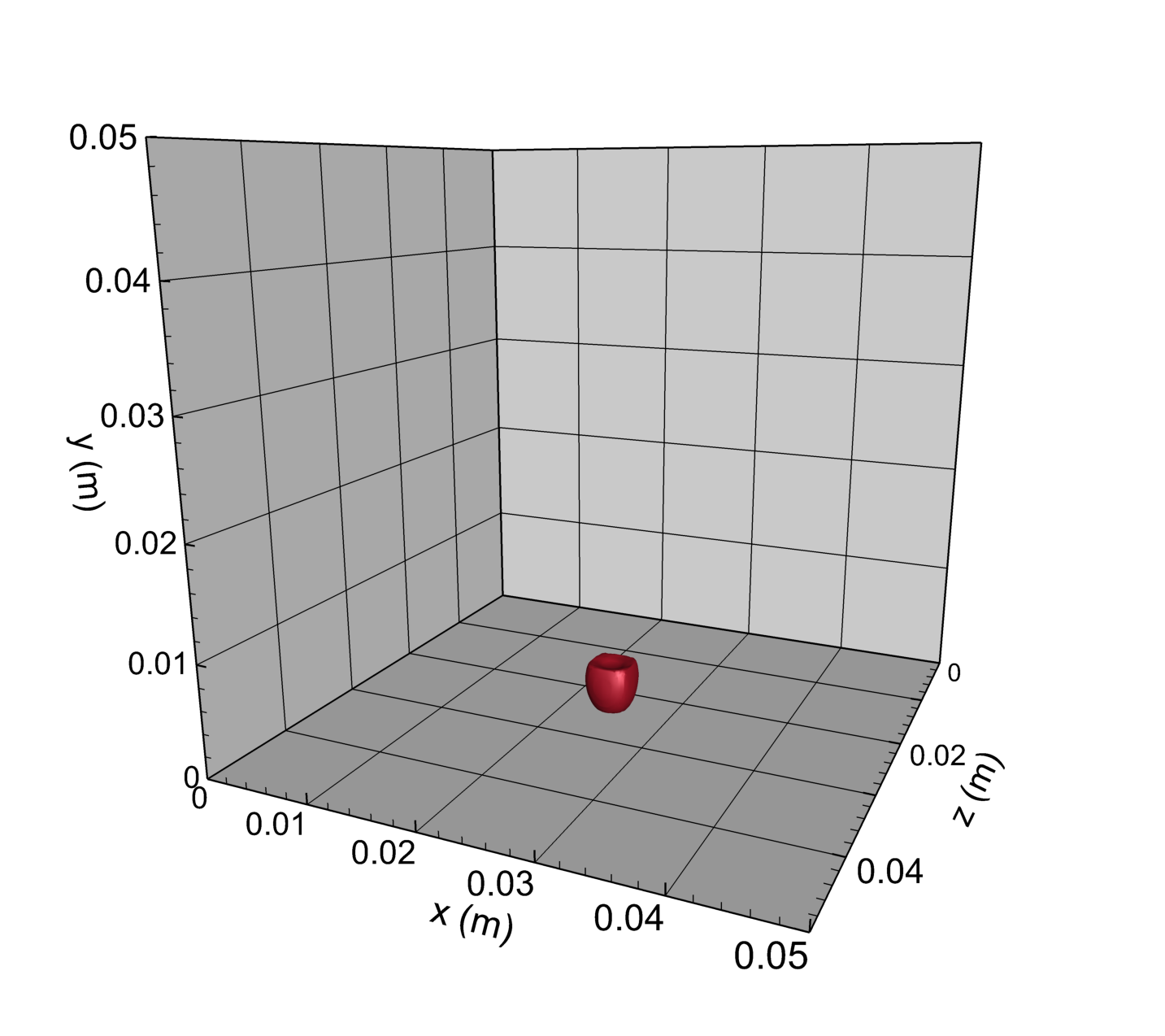}
\caption{\coto jet A: Computational domain including the steady-state
  $\Mach=2.5$ isocontour in red. The inlet is in the middle of the $x$-$z$
  plane.}
\label{fig:co2jet-domain}
\end{figure}

The computational domain, shown in Figure~\ref{fig:co2jet-domain}, is a
cube of edge length \SI{5}{\centi\metre}, divided into an equidistant
grid. The hole geometry is represented by a Cartesian approximation and the
inflow condition is described in Section~\ref{sec:bernoulli}. At the inlet
edge, the boundary condition outside the hole is a no-slip wall. The other
boundaries are governed by the NSCBC (see Section \ref{sec:openbc})
employing a far-field pressure of \SI{1}{\bar} and an $\alpha$-coefficient
of 10.0. Here, the jet flow is in the $y$ direction.

Two grids were employed, with $200^3$ and $400^3$ cells. In the latter
case, the inlet hole had eight cells across the diameter. The computations were performed using the fourth-order accurate quadrature
rule (see Section~\ref{sec:spatdisc}) with the WENO scheme and $C_\cfl =
0.333$. In the WENO scheme, the internal energy, density and velocity were
used as reconstruction variables, and this combination was found to work
well.

\subsubsection{Results}
\begin{figure*}[tbp]
  \centering
  \begin{subfigure}[b]{0.49\linewidth}
    \centering
    \includegraphics[width=\linewidth]{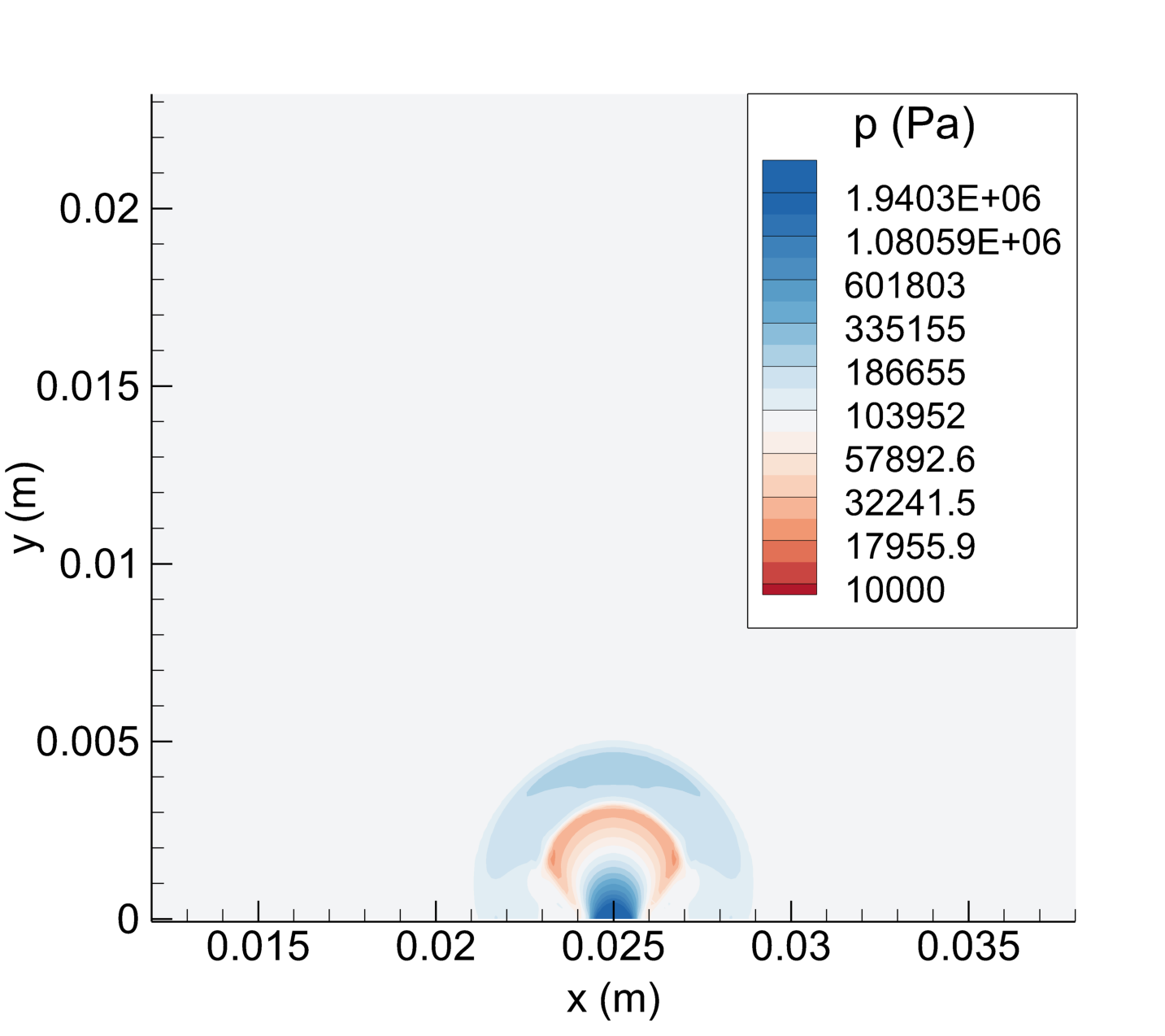}
    \caption{$t=\SI{10}{\micro\second}$.}
    \label{fig:co2jet-p-10mus}
  \end{subfigure}
  \hfill
  \begin{subfigure}[b]{0.49\linewidth}
    \centering
    \includegraphics[width=\linewidth]{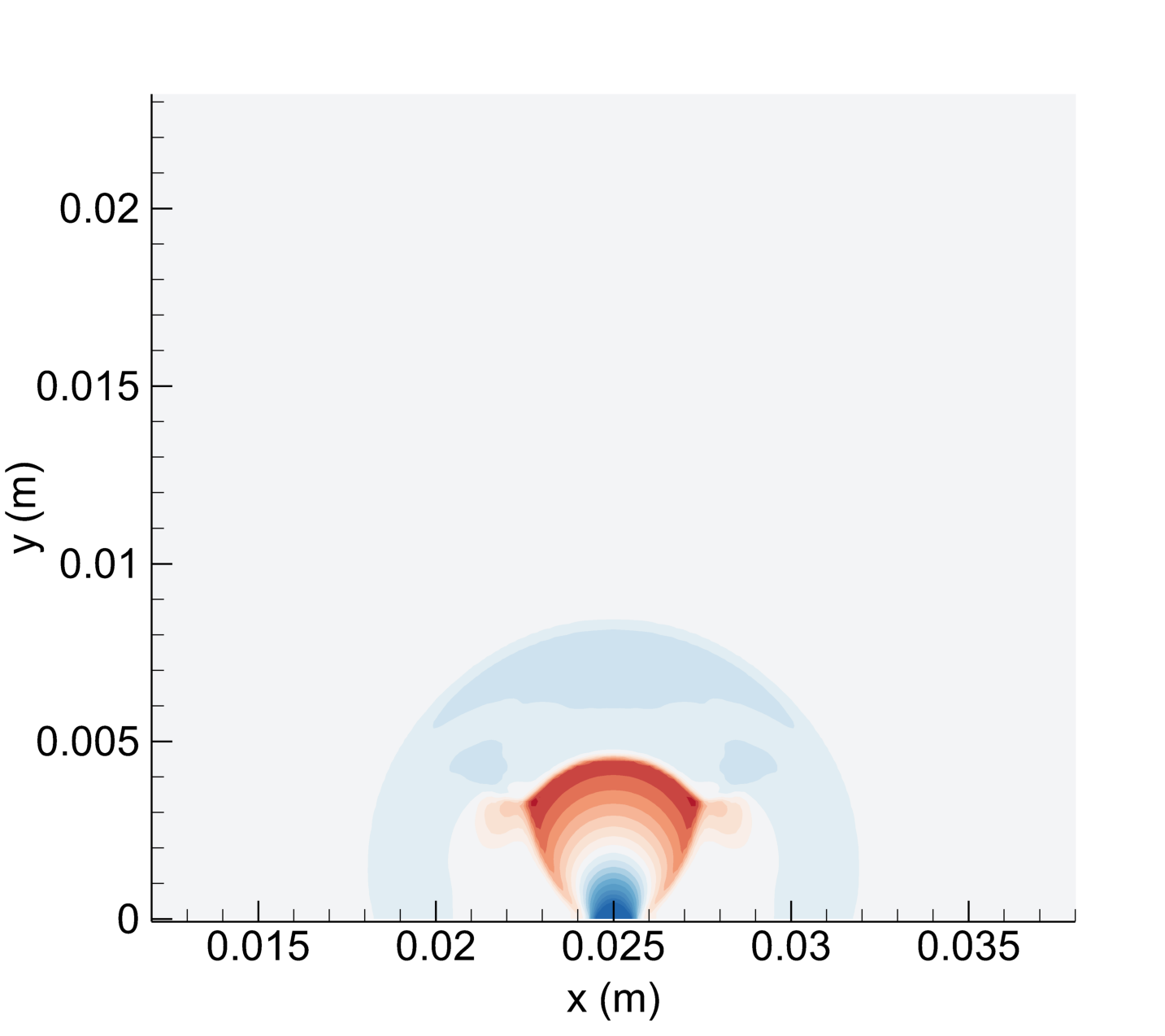}
    \caption{$t=\SI{20}{\micro\second}$.}
    \label{fig:co2jet-p-20mus}
  \end{subfigure}\\
  \begin{subfigure}[b]{0.49\linewidth}
    \centering
    \includegraphics[width=\linewidth]{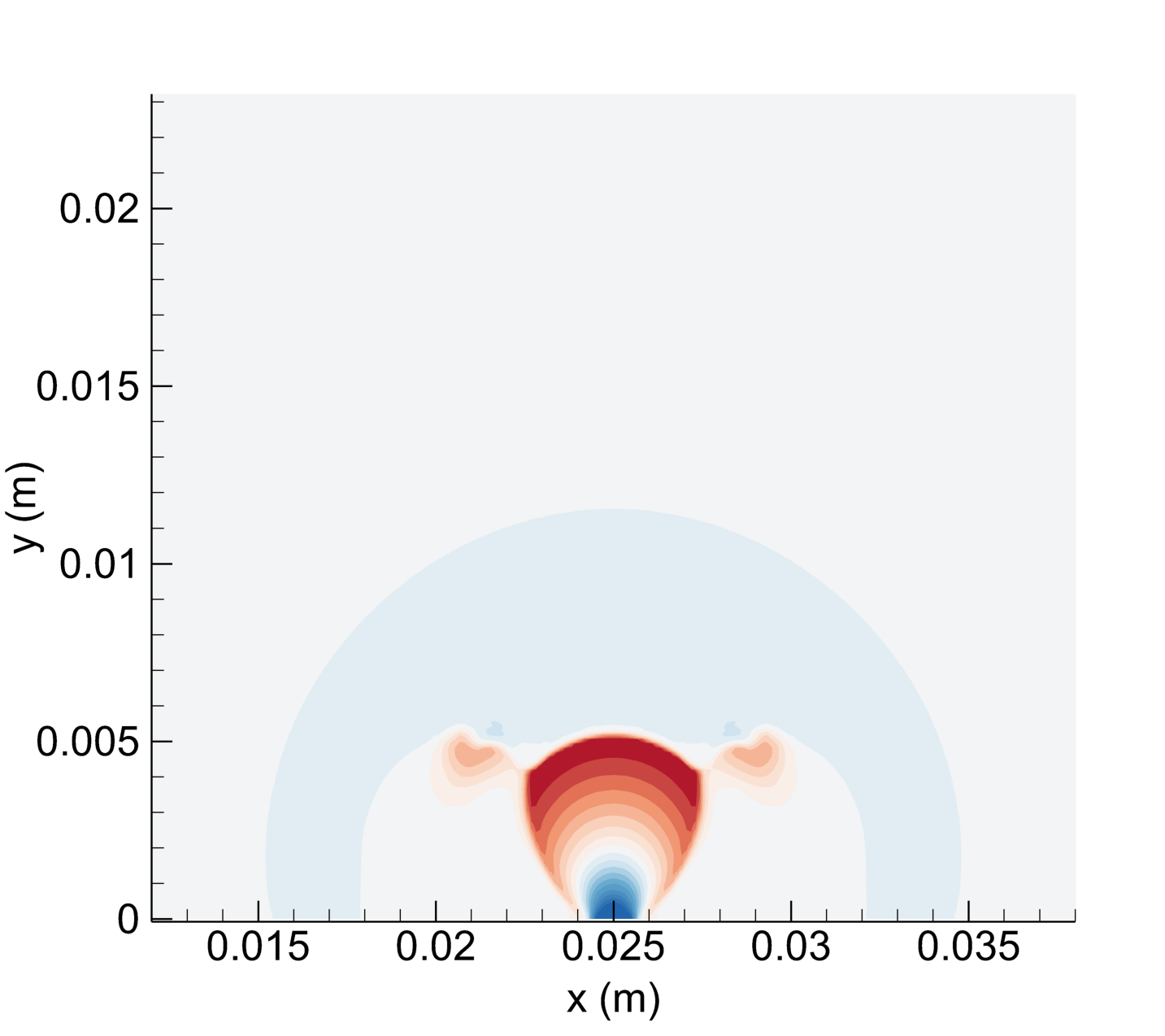}
    \caption{$t=\SI{30}{\micro\second}$.}
    \label{fig:co2jet-p-30mus}
  \end{subfigure}\hfill
  \begin{subfigure}[b]{0.49\linewidth}
    \centering
    \includegraphics[width=\linewidth]{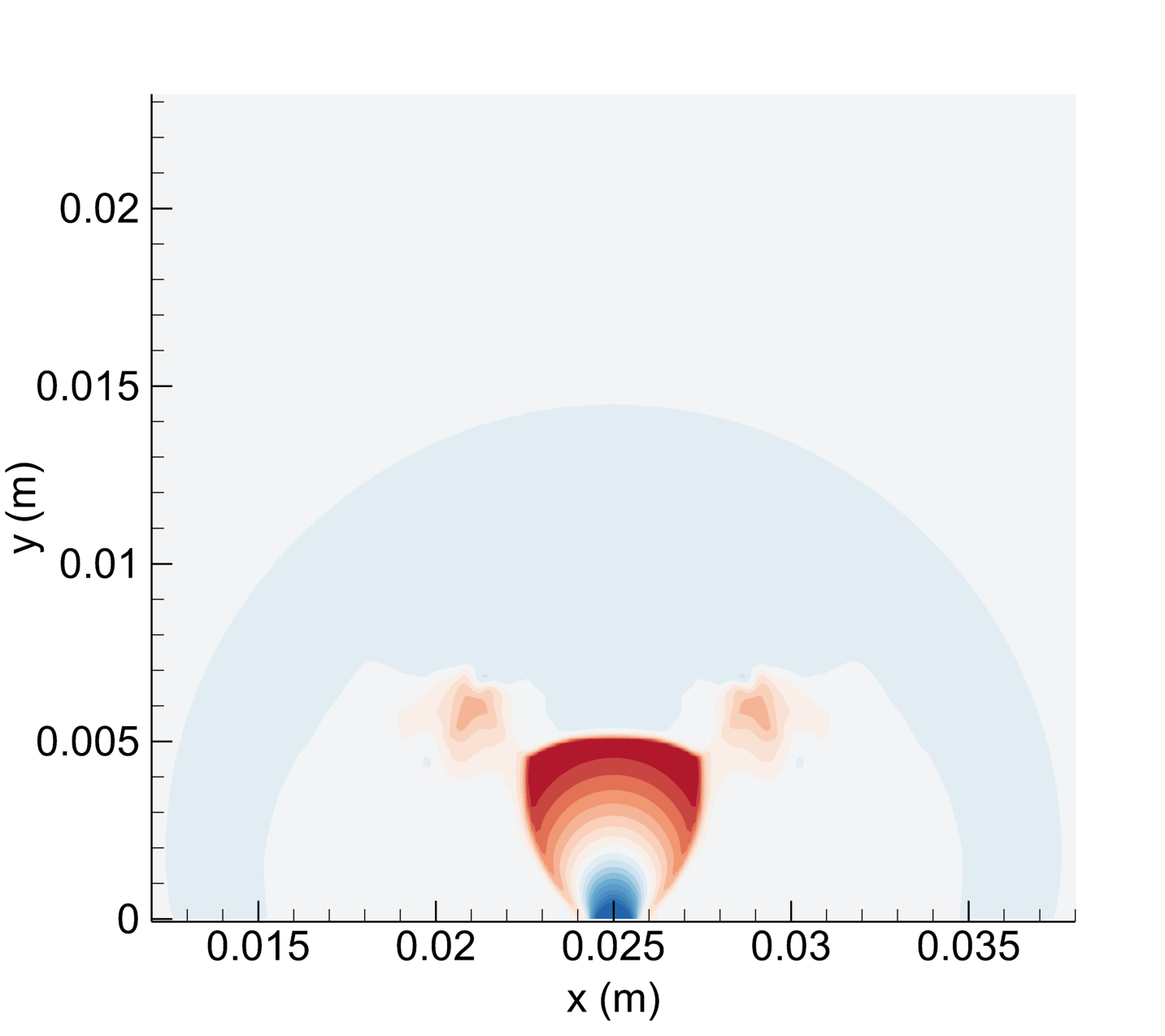}
    \caption{$t=\SI{40}{\micro\second}$.}
    \label{fig:co2jet-p-40mustemp}
  \end{subfigure}
  \caption{\coto jet A: Pressure contours of the developing shock. Plane through
    $z=\SI{0.025}{\metre}$. $400^3$ computational cells.}
  \label{fig:co2jet-p-devel}
\end{figure*}

Pressure contours of the developing barrel shock are displayed in
Figure~\ref{fig:co2jet-p-devel}, while Figure~\ref{fig:co2jet-mach} shows
Mach-number contours at $t=\SI{0.5}{\milli\second}$. In the computation,
the state is gas-liquid-solid in a small region close to the nozzle
exit. The gas-solid state is found in a larger region, particularly inside
the barrel shock, but there is also some solid in the zone beyond the
barrel shock. This is illustrated in Figure~\ref{fig:co2jet-mfs}. In the
figure, the solid mass fraction is indicated by the different coloured
contours, while the liquid area, appearing near the inlet, is denoted by
green colour, which is the \SI{1}{\percent} isocontour of the liquid mass
fraction. Although quantitative measurements are not available in
\citet{Li2016}, we expect the present method to underestimate the
post-shock solid mass fraction in this case due to the assumption of full
thermodynamic equilibrium.

Temperature contours are shown in Figure~\ref{fig:co2jet-temp}. It can be
seen that the \coto jet core is at about \SI{-70}{\celsius}, while the
coldest temperature, about \SI{-100}{\celsius}, is appearing right before
the Mach disk, due to the strong expansion.

\begin{figure}
\centering
\includegraphics[width=0.5\linewidth]{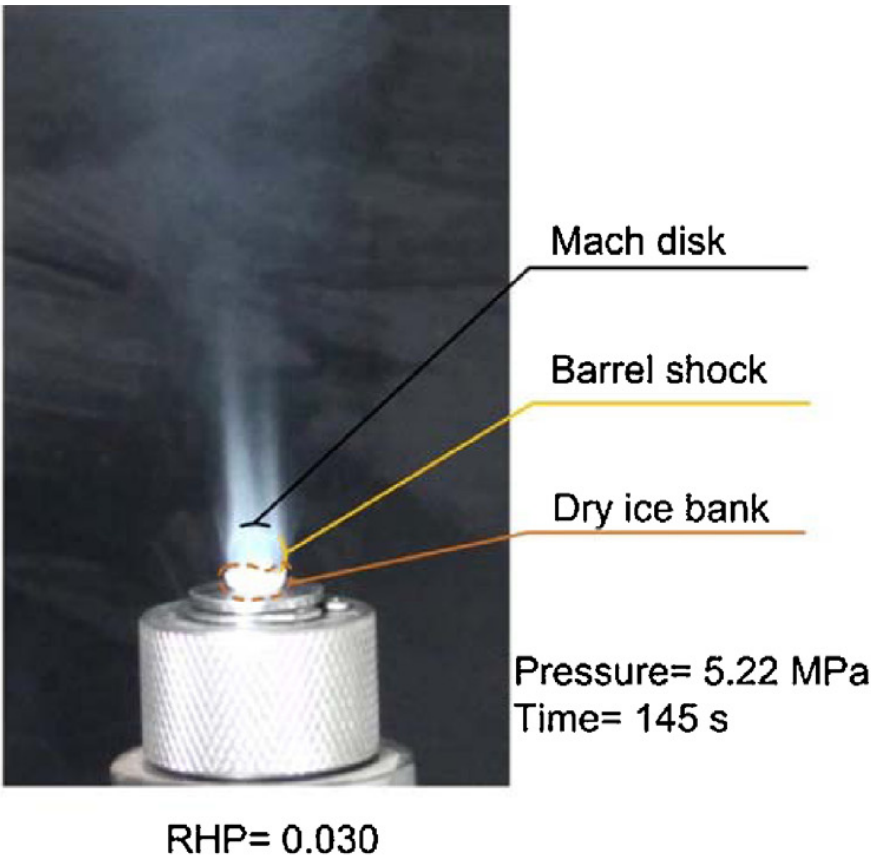}
\caption{Photograph of \coto jet A. Reprinted from Figure~7 in
  \citet{Li2016}, copyright (2016), with permission from Elsevier.}
\label{fig:co2jet-photo}
\end{figure}

\begin{figure*}[tbp]
  \centering
  \begin{subfigure}[b]{0.49\linewidth}
    \centering
    \includegraphics[width=\linewidth]{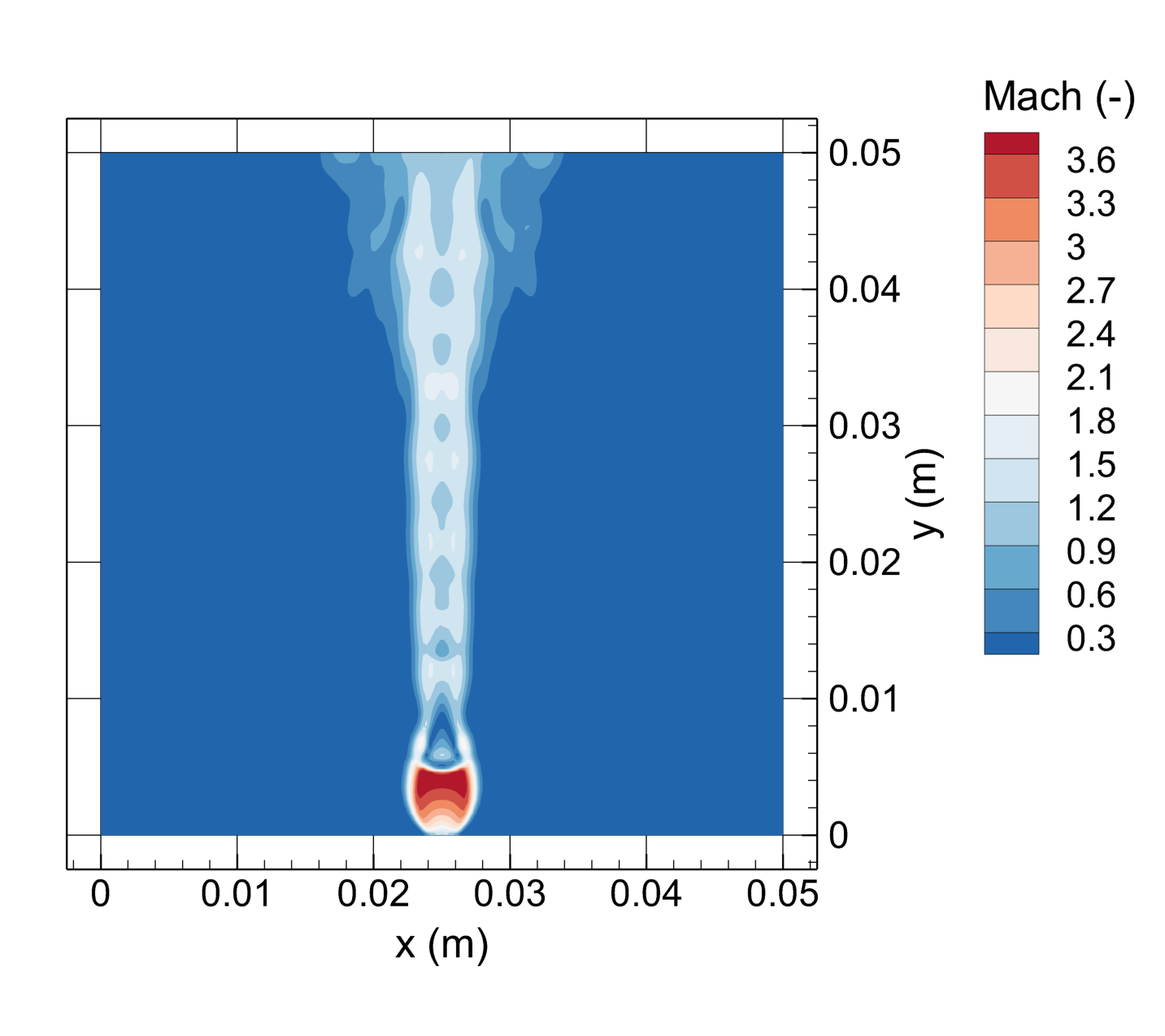}
    \caption{Mach number.}
    \label{fig:co2jet-mach}
  \end{subfigure}
  \hfill
  \begin{subfigure}[b]{0.49\linewidth}
    \centering
    \includegraphics[width=\linewidth]{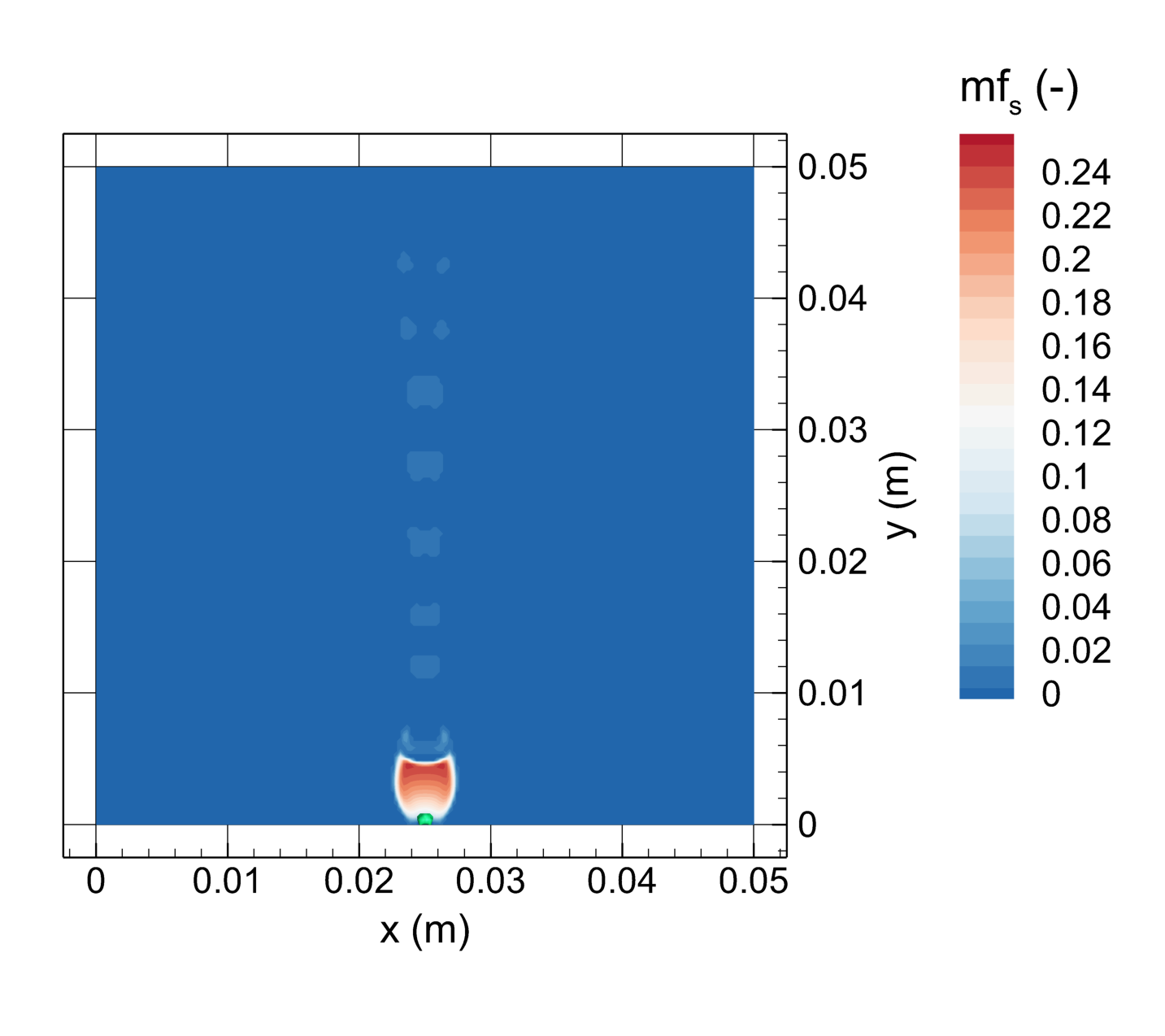}
    \caption{Solid mass fraction. Liquid appears in the green area near the
      inlet.}
    \label{fig:co2jet-mfs}
  \end{subfigure}\\
  \begin{subfigure}[b]{0.49\linewidth}
    \centering
    \includegraphics[width=\linewidth]{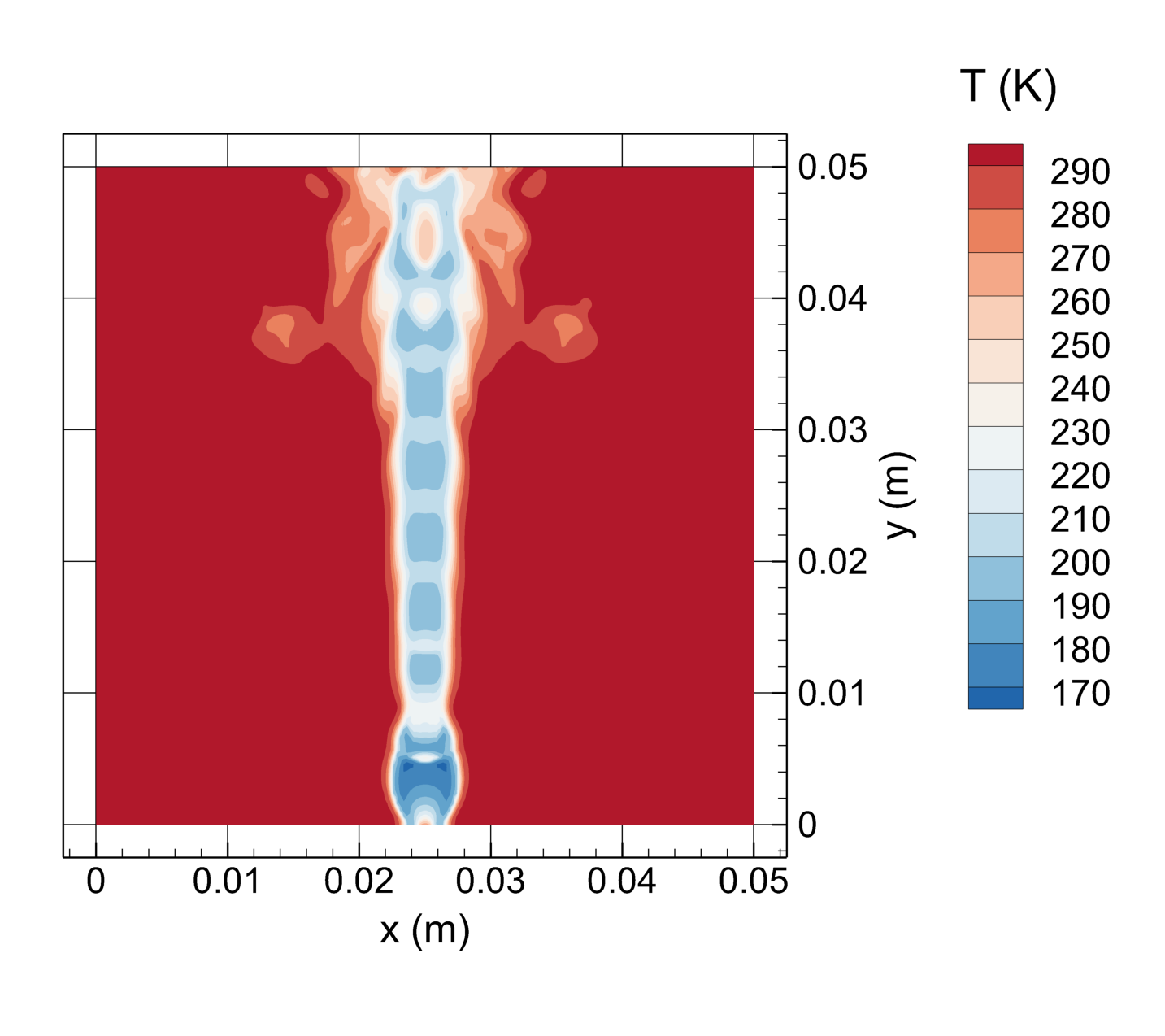}
    \caption{Temperature.}
    \label{fig:co2jet-temp}
  \end{subfigure}\hfill
  \begin{subfigure}[b]{0.49\linewidth}
    \centering
    \includegraphics[width=\linewidth]{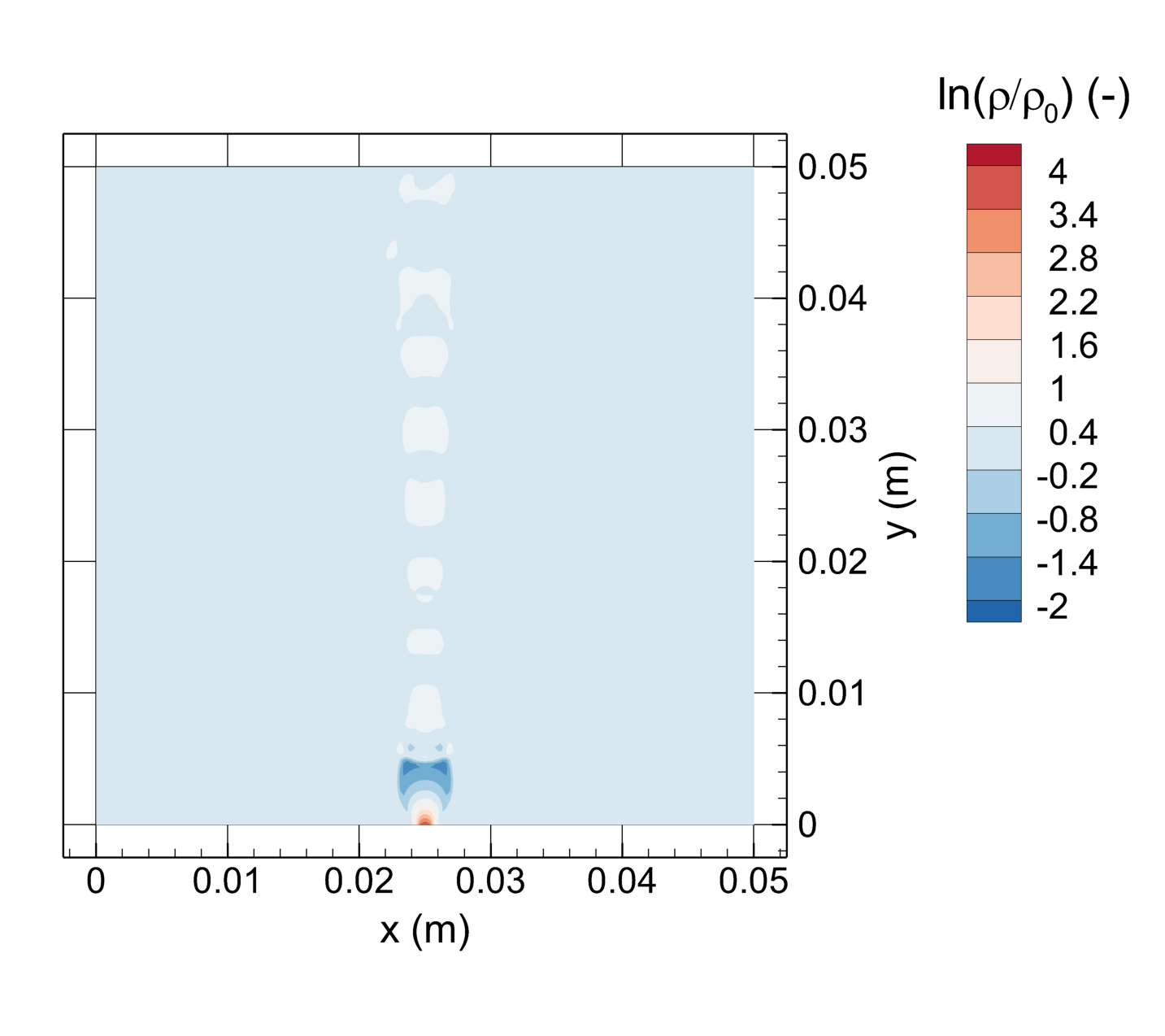}
    \caption{Logarithmic density.
      $\rho_0=\SI{1.82}{\kilogram\per\cubic\metre}$ corresponds to
      the initial state.}
    \label{fig:co2jet-dens}
  \end{subfigure}\hfill
  \caption{\coto jet A: Snapshots at \SI{0.5}{\milli\second}, where the
    barrel shock is in a steady state. Plane through
    $z=\SI{0.025}{\metre}$. $200^3$ computational cells.}
  \label{fig:co2jet-steady}
\end{figure*}

\begin{figure}
\centering
  \begin{subfigure}[b]{0.49\linewidth}
    \centering
    \includegraphics[width=0.9\linewidth]{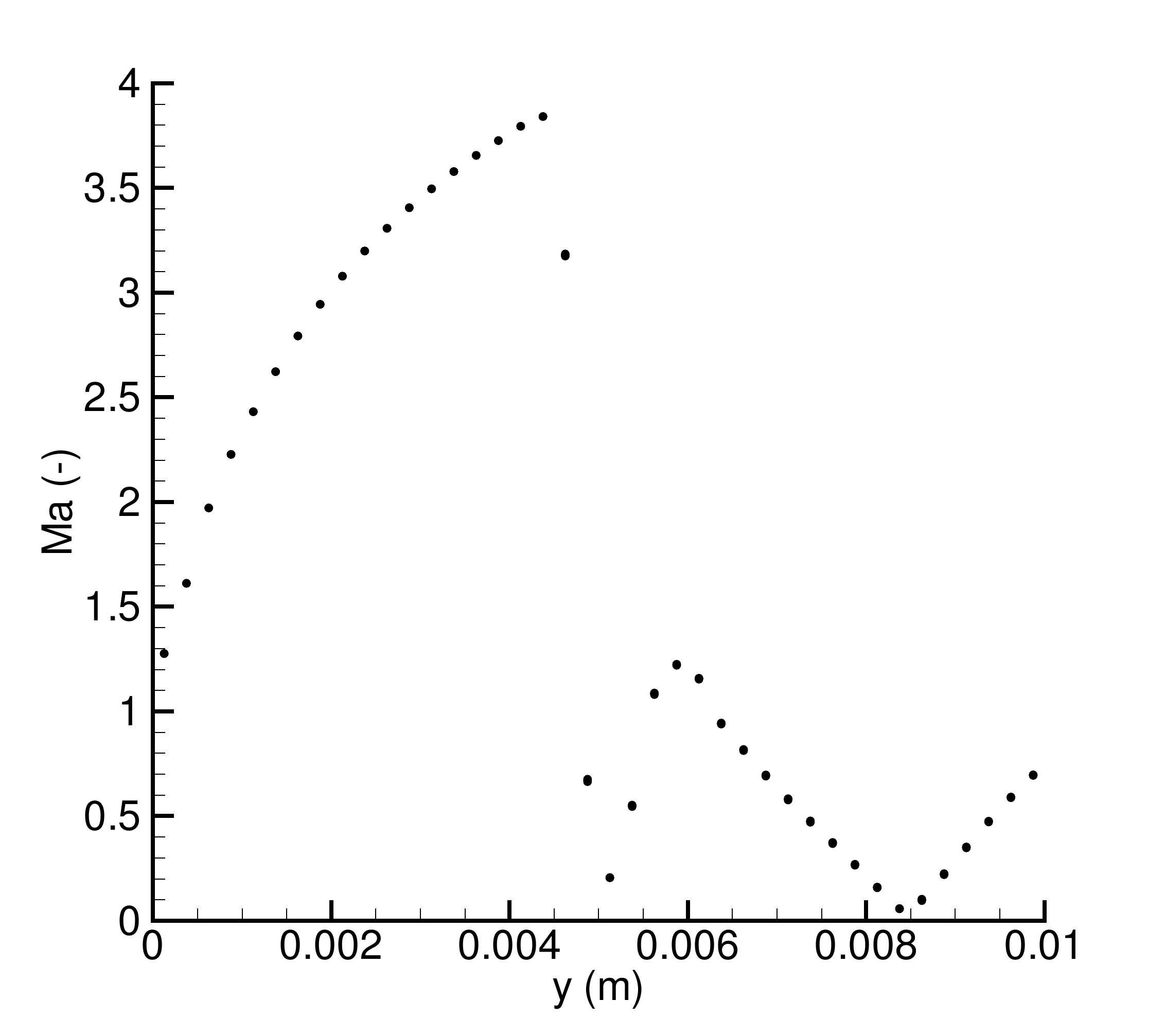}
    \caption{Mach number.}
    \label{fig:co2jet-mach-xy}
  \end{subfigure}\hfill
  \begin{subfigure}[b]{0.49\linewidth}
    \centering
    \includegraphics[width=0.9\linewidth]{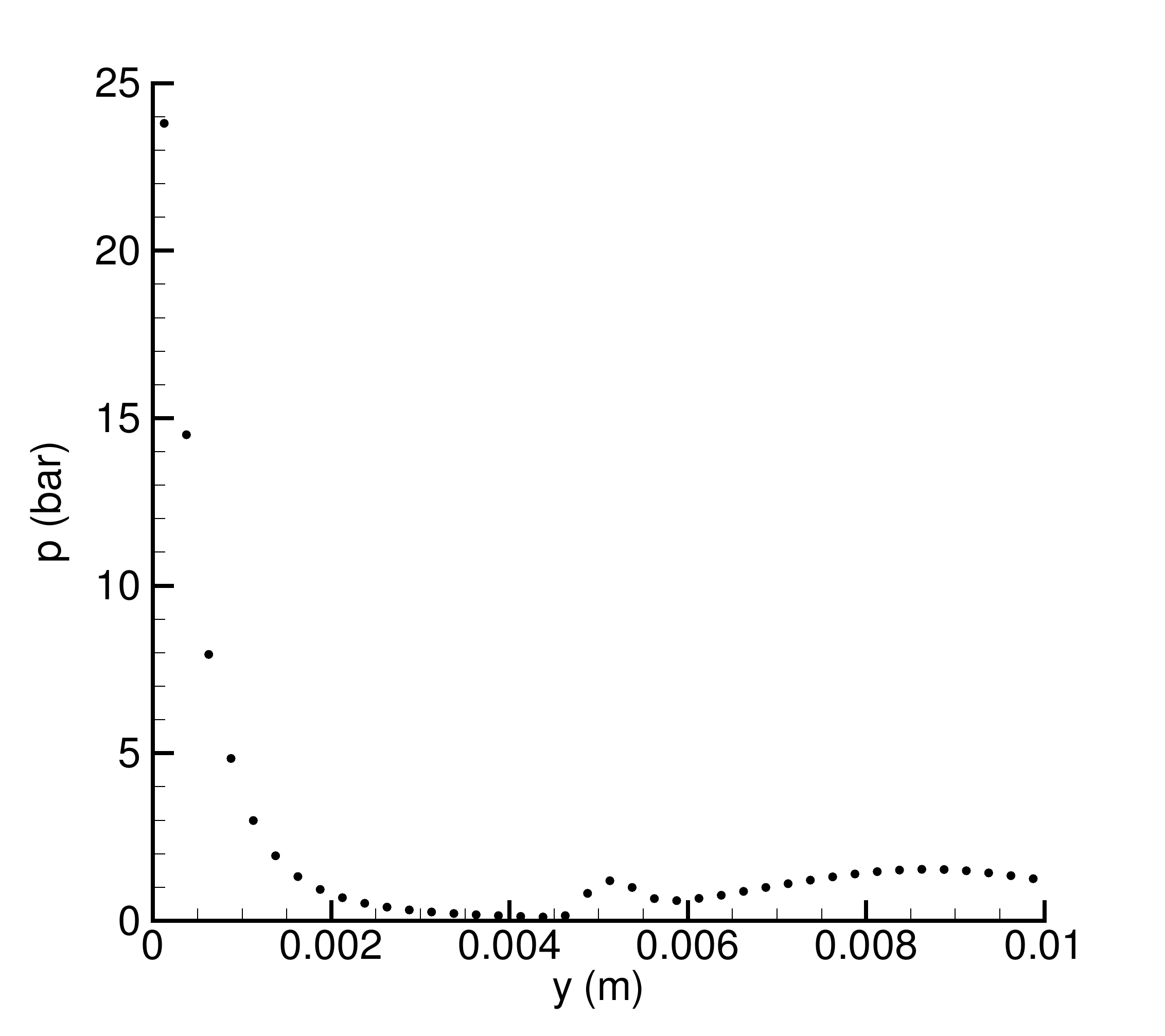}
    \caption{Pressure.}
    \label{fig:co2jet-p-xy}
  \end{subfigure}\\
\caption{\coto jet A: Mach number and pressure plotted along the jet centreline at \SI{0.5}{\milli\second}.}
\label{fig:co2jet-xy}
\end{figure}

In Figure~\ref{fig:co2jet-xy}, the Mach number and the pressure are plotted
along the jet centreline. It can be seen that the position of the Mach
disk is $l\ped{MD}=\SI{4.8}{\milli\metre}$.  This position can also be
estimated by the correlation recommended by \citet{Franquet2015}:
\begin{equation}
  \label{eq:LMD}
  l\ped{MD} = d\ped{e}\sqrt{\frac{p/p_0}{2.4}} \approx \SI{4.7}{\milli\metre}.
\end{equation}
This is in good agreement with the present result.

An accurate position of the Mach disk cannot be obtained from the
photograph in Figure~\ref{fig:co2jet-photo} or the description in
\citet{Li2016}. However, we estimate that the position of the Mach disk is
roughly at half a centimetre in that photograph. Thus there is good
agreement between the photograph and the present result. Further, in
Figure~\ref{fig:co2jet-mach}, it can be seen that the boundary layer around
the jet starts widening at about \SI{4}{\centi\metre}. This is also in
agreement with Figure~\ref{fig:co2jet-photo}. However, in that photograph,
one can see ice at the nozzle exit, labelled ``dry ice bank''. In view of
the high \coto exit velocity, we find it unlikely that dry-ice particles
deposit immediately at the exit. The ice may well be frozen moisture from
the surrounding air.

\subsection{Case B}
Finally, we consider the \coto jet presented as Test B in
\citet{Pursell2012}.

\subsubsection{Case description and computational set-up}
The set-up is analogous to that in Case A, except that the orifice diameter
is four times larger. The inlet condition corresponds to saturated
gas. Pure \coto of temperature $T=\SI{-3.5}{\celsius}$ and pressure
$p=\SI{31.0}{\bar}$ is expanded through a round hole of diameter
$d\ped{e}=\SI{4}{\milli\metre}$. The Reynolds number at the inlet is
\num{2e6}. The initial temperature in the computational domain is assumed
to be $T_0=\SI{20}{\celsius}$.

The computational domain is a cube of edge length \SI{10}{\centi\metre},
divided into an equidistant grid with 200 cells in each direction.

\subsubsection{Results}

\begin{figure}
\centering
\includegraphics[width=0.5\linewidth]{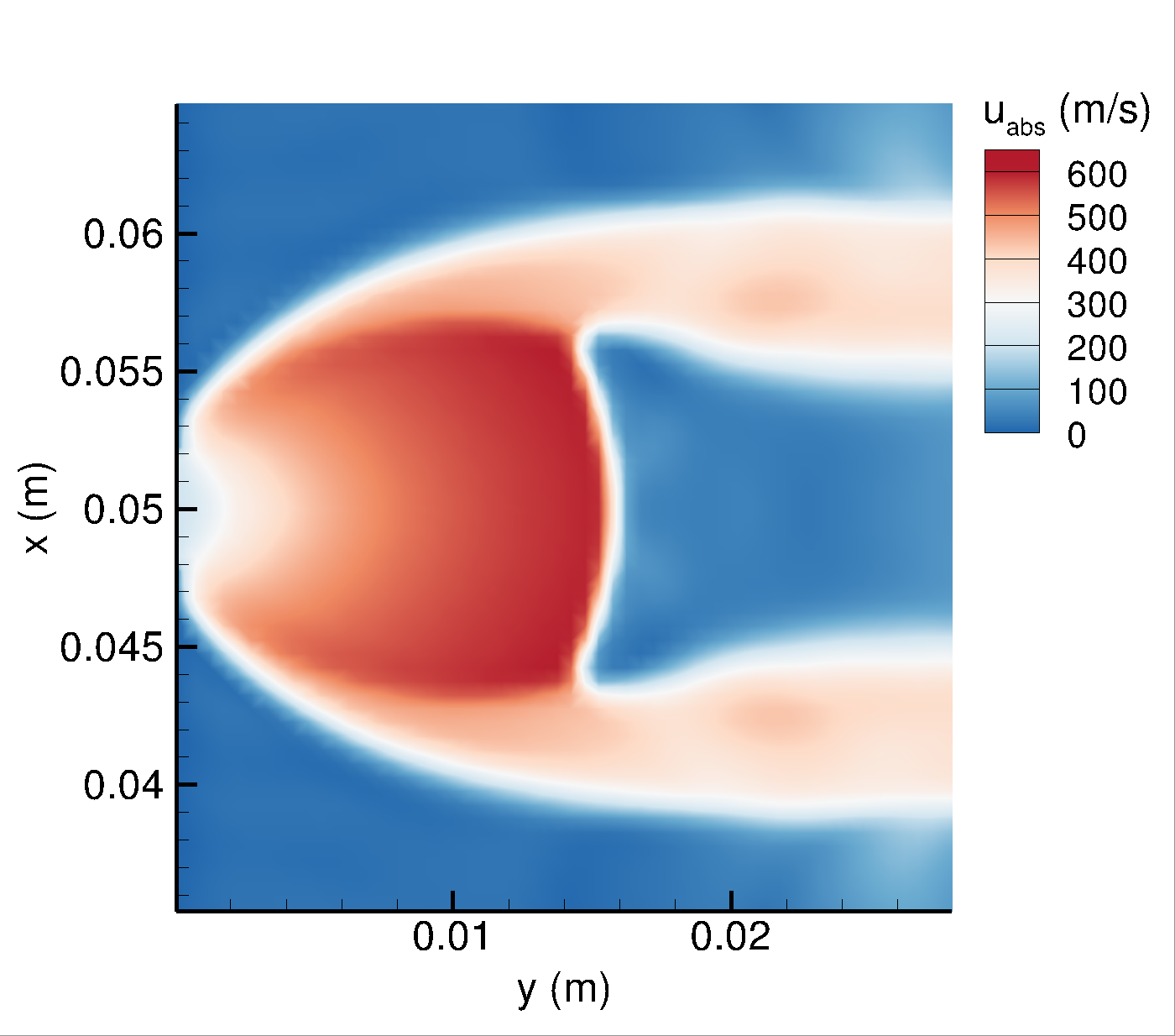}
\caption{\coto jet B: Absolute velocity in a plane through
    $z=\SI{0.025}{\metre}$. Snapshot at $t=\SI{0.28}{\milli\second}$.}
\label{fig:co2jet-uabs}
\end{figure}

Figure~\ref{fig:co2jet-uabs} shows contours of the absolute velocity
plotted at $t=\SI{0.28}{\milli\second}$. The
Mach-disk position is at $\SI{16.1}{\milli\metre}$ and the width of the
structure is $\SI{21.7}{\milli\metre}$. In this case, the
correlation~\eqref{eq:LMD} gives $l\ped{MD}\approx\SI{14.4}{\milli\metre}$
for the Mach-disk position, which is a difference of
$\SI{12}{\percent}$ -- somewhat larger than in Case A.

The experimental values found by \citet{Pursell2012} are
$\SI{17.8}{\milli\metre}$ and $\SI{21.9}{\milli\metre}$, for the Mach-disk
position and the `effective diameter', respectively. Hence, the simulated
Mach-disk position lies between that measured by Pursell and the one given
by the correlation. The agreement is very good between the simulated and
measured effective diameter.  Further, the barrel-shock structure seen in
Figure~\ref{fig:co2jet-uabs} closely corresponds to the photograph given in
Figure~5(b) in \citet{Pursell2012}.

\section{Conclusions}
\label{sec:concl}

We have developed a high-order finite-volume method capable of handling
single-phase, two-phase (gas-liquid) and three-phase (gas-solid-liquid)
flow with discontinuities. The spatial and temporal discretization is
similar to that of \citet{Coralic2014}, employing a fifth-order WENO scheme
and a third-order strong-stability-preserving Runge--Kutta method. In the
present case, however, the model formulation is based on the homogeneous
equilibrium model. The fluid phase behaviour is calculated using a suitable
EOS and assuming full local thermodynamic equilibrium. This
approach requires special care in the implementation of the flash
algorithms required to translate the state variables into primitive
variables.

The method has been validated using various benchmark cases, showing robust
behaviour and high-order convergence for smooth single- and two-phase flow.
Further, the calculation of a turbulent air jet was validated by comparing
with data from \citet{Deo07}. Finally, the method was employed to calculate
a highly underexpanded \coto jet exhibiting phase transition and
three-phase flow. To this end, the \citet{Span1996} reference EOS was
employed. The method was able to robustly and accurately capture the
complex and intertwined thermo- and fluid dynamics. The shape and
dimensions of the barrel shock closely corresponded with the observations
of \citet{Pursell2012}.  The position of the Mach disk was correctly
predicted with reference to the correlation recommended by
\citet{Franquet2015}, and good agreement was obtained with the photograph
of \citet{Li2016}.

We intend to further develop the method and apply it for the calculation of
complex \coto flows occurring in jets, process equipment and pipelines or wells.

\section*{Acknowledgments}
This work was carried out in the 3D Multifluid Flow project funded by
the Research Council of Norway through the basic grant to SINTEF
Energy Research. Some of the computations were performed thanks to a
resource allocation at the Notur high-performance computing
infrastructure (NN9432K).

\appendix
\section{Characteristic decomposition for general EOS}
\label{sec:chars}
\citet{Coralic2014} perform reconstruction in the local characteristic
variables. They calculate these by multiplying their vector of primitive
variables with locally valid transformation matrices. The matrices they give are
valid for the stiffened-gas EOS. Since the ideal-gas EOS is a special case of
the stiffened-gas EOS, they are also valid for ideal gas. Here we present
transformation matrices that can be used with a general EOSs.

We motivate the procedure for computing the local characteristic variables by
first rewriting the system \eqref{eq:hem} into a quasi-linear form in
terms of a set of variables $\vec{R}$,
\begin{equation}
  \label{eq:hem_primitive}
  \partial_t \vec{R} + \vec{A} \partial_x \vec{R}
  + \vec{B} \partial_y \vec{R}
  + \vec{C} \partial_z \vec{R} = \vec{0}.
\end{equation}
Herein, $\vec{A} = \partial_{\vec{R}} \vec{F}$, $\vec{B}
= \partial_{\vec{R}} \vec{G}$ and $\vec{C} = \partial_{\vec{R}} \vec{H}$
are the Jacobian matrices. Like \citet{Coralic2014}, we have omitted the
diffusive fluxes and source terms as we are interested in the
characteristics of the advective equation system. The vector $\vec{R}$ is
\begin{equation}
  \vec{R} = \left[ \rho, u_x, u_y, u_z, s \right]^\transp,
\end{equation}
and the Jacobian matrices are
\begin{equation}
  \vec{A} =
  \begin{bmatrix}
    u_x & \rho & 0 & 0 & 0\\
    \frac{c^2}{\rho} & u_x & 0 & 0 &
    \frac{1}{\rho}\pdc{p}{s}{\rho} \\
    0 & 0 & u_x & 0 & 0 \\
    0 & 0 & 0 & u_x & 0 \\
    0 & 0 & 0 & 0 & u_x \\
  \end{bmatrix},
\end{equation}
\begin{equation}
  \vec{B} =
  \begin{bmatrix}
    u_y & 0 & 0 & 0 & 0\\
    0 & u_y & 0 & 0 & 0 \\
    \frac{c^2}{\rho} & 0 & u_y & 0 &
    \frac{1}{\rho}\pdc{p}{s}{\rho} \\
    0 & 0 & 0 & u_y & 0 \\
    0 & 0 & 0 & 0 & u_y \\
  \end{bmatrix},
\end{equation}
and
\begin{equation}
  \vec{C} =
  \begin{bmatrix}
    u_z & 0 & 0 & 0 & 0\\
    0 & u_z & 0 & 0 & 0 \\
    0 & 0 & u_z & 0 & 0 \\
    \frac{c^2}{\rho} & 0 & 0 & u_z &
    \frac{1}{\rho}\pdc{p}{s}{\rho} \\
    0 & 0 & 0 & 0 & u_z \\
  \end{bmatrix},
\end{equation}
where $c$ is the speed of sound.

The equation system \eqref{eq:hem_primitive}, has one set of
characteristic variables for each direction $x$, $y$ and $z$. Here we
consider the $x$-direction only, as the treatment of $y$ and $z$ is
analogous. The matrix $\vec{A}$ can be written in terms of a diagonal
decomposition,
\begin{equation}
  \vec{A} = \vec{P} \vec{\Lambda} \vec{P}^{-1}
\end{equation}
where $\vec{\Lambda}$ is the diagonal matrix of eigenvalues and the columns of
$\vec{P}$ are the right eigenvectors of $\vec{A}$.
\begin{equation}
  \vec{\Lambda} =
  \begin{bmatrix}
    u_x & 0 & 0 & 0     & 0     \\
    0 & u_x & 0 & 0     & 0     \\
    0 & 0 & u_x & 0     & 0     \\
    0 & 0 & 0 & u_x - c & 0     \\
    0 & 0 & 0 & 0       & u_x + c \\
  \end{bmatrix}
\end{equation}

\begin{equation}
  \vec{P} =
  \begin{bmatrix}
    0 & 0 & -\frac{1}{c^2}\pdc{p}{s}{\rho} & -\frac{\rho}{c} & \frac{\rho}{c}
    \\
    0 & 0 & 0 & 1 & 1 \\
    1 & 0 & 0 & 0 & 0 \\
    0 & 1 & 0 & 0 & 0 \\
    0 & 0 & 1 & 0 & 0
  \end{bmatrix}
\end{equation}

\begin{equation}
  \vec{P}^{-1} =
  \begin{bmatrix}
    0 & 0 & 1 & 0 & 0 \\
    0 & 0 & 0 & 1 & 0 \\
    0 & 0 & 0 & 0 & 1 \\
    -\frac{c}{2\rho} & \frac{1}{2} & 0 & 0 &
    -\frac{1}{2\rho c}\pdc{p}{s}{\rho} \\
    \frac{c}{2\rho} & \frac{1}{2} & 0 & 0 &
    \frac{1}{2\rho c}\pdc{p}{s}{\rho}
  \end{bmatrix}
\end{equation}
If we now assume that the matrices $\vec{P}$ and $\vec{P}^{-1}$ and locally
frozen in time and space, and thus commute with the differential operators, we
can write the $x$-direction part of \eqref{eq:hem_primitive} as
\begin{equation}
  \label{eq:hem_char_x}
  \partial_t \vec{X} = \vec{\Lambda} \partial_x \vec{X},
\end{equation}
where
\begin{equation}
  \vec{X} = \vec{P}^{-1} \vec{R}.
\end{equation}
The equation system \eqref{eq:hem_char_x} has the form of a decoupled
system of advection equations and is locally valid to to the extent
that the local temporal and spatial variations in $\vec{R}$ are
small. Thus $\vec{X}$ is a local approximation to the characteristic
variables for advection in the $x$-direction.

We apply this approximation when performing reconstruction in the local
characteristic variables. Before reconstruction to the cell edge quadrature
points at $i+1/2,j,k$, we calculate the characteristic variables in all cells
$\ell, m, n$ in the stencil,
\begin{equation}
  \vec{X}_{\ell,m,n} = \vec{P}^{-1}_{i+1/2,j,k} \vec{R}_{\ell,m,n}.
\end{equation}
using the same projection matrix $\vec{P}^{-1}_{i+1/2,j,k}$. The projection
matrix is calculated using a simple arithmetic mean of the fluid state at
$i,j,k$ and $i+1,j,k$.

After reconstruction, we have the characteristic variables at the quadrature
points on the left and right side of the cell edge $i+1/2,j,k$. The variables
$\vec{R}$ at these points are obtained by multiplying with the inverse
projection matrix,
\begin{linenomath}
\begin{align}
  \vec{R}^\leftedge_{i+1/2,j_\ell,k_m}
  &= \vec{P}_{i+1/2,j,k} \vec{X}^\leftedge_{i+1/2,j_\ell,k_m}, \\
  \vec{R}^\rightedge_{i+1/2,j_\ell,k_m}
  &= \vec{P}_{i+1/2,j,k} \vec{X}^\rightedge_{i+1/2,j_\ell,k_m}.
\end{align}
\end{linenomath}

\phantomsection
\addcontentsline{toc}{section}{References}
\providecommand{\coto}{{CO$_2$}} \providecommand{\soto}{{SO$_2$}}
  \providecommand{\nto}{{N$_2$}} \providecommand{\hto}{{H$_2$}}
  \providecommand{\oto}{{O$_2$}} \providecommand{\chf}{{CH$_4$}}
  \providecommand{\natosof}{{Na$_2$SO$_4$}}
  \providecommand{\nahcoth}{{Na$_2$HCO$_3$}} \providecommand{\htoo}{{H$_2$O}}

\end{document}